# A View of How Language Models Will Transform Law

*Frank Fagan*[*]

*Abstract*

This Article considers the influence of Large Language Models (LLMs) on legal practice and the legal services industry. In the near term, LLMs will spur new legal work. Lawyers will be called upon to help litigate new questions over property rights in data, language model output, and lawyer-engineered prompts. Lawyers will additionally help judges decide what to do about new forms of torts, including legal malpractice, enabled by the casual and lightly supervised use of large language models. As legal rules governing the use of generative A.I. begin to clarify and settle, and as the technology fully matures, future lawyers faced with routine work will engage language models to save time and costs. Consequently, legal tasks will take less time to complete, and language models will enhance lawyer productivity.

While most commentators have focused exclusively on how LLMs will transform day-to-day law practice, a substantial structural change could be afoot within the legal sector as a whole. Large increases in productivity and attendant cost savings could encourage law firms and corporate legal departments to develop large language models in-house. A ten percent increase in attorney productivity would encourage an average sized "Big Law" firm to reduce its associate headcount by 300 to 400 lawyers. This represents cost savings of $60 to $120 million—more than enough to pay for the development of a specialized LLM. Consider a senior partner who relies heavily on a proprietary language model to service a client. If the model is owned and controlled by the firm, then clients will be more strongly tied to the firm and may wish to remain there—even if the partner departs. Generative A.I. thus portends a shift in the balance of power between partners and firms. To what extent, it remains to be seen.

Eventually, LLMs will push lawyers into highly specialized and nuanced roles. After fully mature LLMs arrive, the lawyer will continue to play a central role in legal practice, but only in non-routine legal tasks. These tasks will primarily involve value judgments, such as the development of precedent or its reversal, or the allocation of property and other scarce resources. This new mix of lawyer-machine labor, where machines primarily carry out routine legal tasks, and lawyers handle the non-routine, will give rise to a growing demand for lawyers who can exercise good judgment and empathize with the winners and losers of social change. Overall,

---

[*] Associate Professor, South Texas College of Law Houston; Research Associate, EDHEC Augmented Law Institute (France). Comments welcome: ffagan@stcl.edu. The author thanks Monica Ortale for research assistance and Tom Hogan and Michael Madison for comments.

*the Article suggests a possible future where there are fewer lawyers and greater consolidation of the legal sector.*

## INTRODUCTION

In a widely acclaimed article published in *The New Yorker* this year, science fiction author Ted Chiang analogized ChatGPT to a "blurry JPEG of the web."[1] Chiang criticized ChatGPT, and large language models[2] generally, for paraphrasing the internet.[3] Better to have quotations and lossless data than rewordings and fuzzy artifacts. An anthropologist or archeologist can surely sympathize. It is difficult to reconstruct a skeleton from a jawbone or The Palace of Knossos from a handful of columns. It is far better to unearth an entire carcass or complete wall. The excitement surrounding ChatGPT, says Chiang, is the result of an illusion.[4] By removing a few words and rephrasing others, generative AI tricks us into thinking it has created something novel on its own. It is as if an archeologist, once tasked with unearthing the hidden past, has now become a builder of new things. He removes a column from a dig site, perhaps rearranges a few others, and then presents the transformation as a wall built from scratch. If the analogy holds, then ChatGPT is a fraud, or at least a skilled conjurer.

---

[1] *See* Ted Chiang, *ChatGPT Is a Blurry JPEG of the Web*, THE NEW YORKER (Feb. 9, 2023), https://www.newyorker.com/tech/annals-of-technology/chatgpt-is-a-blurry-jpeg-of-the-web.

In apparent support of Chiang's thesis, ChatGPT recently conjured several illusions[5] in the Southern District of New York.[6] A lawyer, aided by ChatGPT,

---

Noam Chomsky, along with Ian Roberts and Jeffrey Wattumull, has offered a similar critique, describing ChatGPT as a "lumbering statistical engine for pattern matching, gorging on hundreds of terabytes of data and extrapolating the most likely conversational response or most probable answer to a scientific question." Noam Chomsky, Ian Roberts & Jeffrey Wattumull, *The False Promise of ChatGPT*, N.Y. Times (Mar. 8, 2023), https://www.nytimes.com/2023/03/08/opinion/noam-chomsky-chatgpt-ai.html.

[2] This Article uses interchangeably the terms "large language model", "Chat-GPT", and "generative artificial intelligence." Artificial intelligence is a general term that refers to the transfer of tasks from humans to computers. *See* RAYMOND KURZWEIL, THE AGE OF INTELLIGENT MACHINES 2 (1990). When artificial intelligence generates content, it is be termed "generative artificial intelligence." *See* McKinsey & Company, *What Is Generative AI?*, MCKINSEY.COM (Jan. 19, 2023) https://www.mckinsey.com/featured-insights/mckinsey-explainers/what-is-generative-ai#/.
Large language models (LLMs) are a specific type of generative artificial intelligence that generate language in response to user input. *See* Gary Drenik, *Large Language Models Will Define Artificial Intelligence*, FORBES, Jan. 11, 2023, https://www.forbes.com/sites/garydrenik/2023/01/11/large-language-models-will-define-artificial-intelligence/?sh=4c9d4971b60f. Chat-GPT is a specific large language model that has been in development for the past decade. *See* Bernard Marr, *A Short History of ChatGPT: How We Got To Where We Are Today*, FORBES, May 19, 2023, https://www.forbes.com/sites/bernardmarr/2023/05/19/a-short-history-of-chatgpt-how-we-got-to-where-we-are-today/?sh=5a53718d674f. It is currently in its fourth major iteration, "Chat-GPT-4." *Id.*

[3] *See* Chiang, *supra* note 1 (asking rhetorically "[w]hat use is there in having something that rephrases the Web" when the actual Web is readily available).

[4] *Id.* (noting that "rephras[ing] material from the Web instead of quoting it word for word makes [ChatGPT] seem like a student expressing ideas in her own words, rather than simply regurgitating what she's read; it creates the illusion that ChatGPT understands the material.").

[5] Computer scientists use the term "hallucination" to describe mistakes and errors made by ChatGPT and other "large language models". *See* Ziwei Ji et al., *Survey of Hallucination in Natural Language Generation*, 55 ACM COMPUTING SURVEY 1-2 (2022) (surveying approaches to measuring and diagnosing the hallucination problem in large language models). In a recent interview given at Singapore Management University, Sam Altman, the CEO of OpenAI (the firm that produces

submitted a brief that contained artificial case law.[7] The lawyer did not check to see if the computer-generated citations were genuine.[8] This episode, while certainly suggestive of Chiang's blurred image, does not necessarily imply that ChatGPT constructs false arguments. It is true that the brief referenced fake case law for its quotations, but the judge observed that most of the quoted material was accurate and genuine.[9] ChatGPT simply used the wrong case names. On this point, lawyers, judges, law professors, and students should pause. It is curious that the citations were manifestly fake, while their surrounding text and arguments were simultaneously correct and defensible. Usually, bad references make for bad writing and logic, but it would seem from this example that law and legal writing is different. What can explain the accuracy of legal argument in the presence of false and erroneous legal citations other than the fact that legal writing itself is a blurred image?

   In law, compression is a virtue, not a form of deceit or an act of fraud.[10] Lawyers distill rules from patterns observed in cases and other contexts and leverage the familiar. They design persuasive briefs by borrowing and discarding legal statements from prior decisions.[11] Even if the lawyer wished to short-circuit this process, she can never comprehensively copy an earlier application of law to facts because each factual scenario presented by a dispute is unique.[12] Legal writing is routine and programmatic, but with a creative purpose. It is an exercise in artful compression and rearrangement.

---

ChatGPT) asserted that hallucinations will be eliminated within 1-2 years. *See* Jerrold Soh, Notes from the OpenAI World Tour in Singapore, June 13, 2023, https://www.linkedin.com/pulse/openai-world-tour-singapore-jerrold-soh/.

   [6] *See* Order to Show Cause at 1, Mata v. Avianca, Inc., No. 22-cv-1461 (PKC) (S.D.N.Y. 2023).

   [7] *Id.* (noting "six of the submitted decisions appear to be bogus judicial decisions").

   [8] The lawyer explained that he made a mistake "not confirming the sources provided by ChatGPT." Memorandum of Law by Non-Parties Stephen A. Schwartz and Levidow Lewidow & Oberman P.C. in Response to May 26, 2023 Order to Show Cause at 14, Mata v. Avianca, Inc., No. 22-cv-1461 (PKC) (S.D.N.Y. 2023). He "thought he was using a new, state-of-the-art search engine, not an artificial language program." *Id.* at 18.

   [9] Order to Show Cause at 2, Mata v. Avianca, Inc., No. 22-cv-1461 (PKC) (S.D.N.Y. 2023).

   [10] And if one party attempts to compress too much in order to deceptively characterize their argument, the other party is expected to expose them through our adversarial process. *See* Robert Cooter & Daniel Rubinfeld, *Economic Analyses of Legal Disputes and Their Resolution*, 27 J. ECON. LIT. 1067, 1072 (1989) (analyzing party behavior at a trial and noting that a party may be encouraged to invest more in presenting her side of the facts to a court when the other party presents a different view).

   [11] *See, e.g.*, EDWARD H. LEVI, AN INTRODUCTION TO LEGAL REASONING 3 (2013) (noting that legal reasoning involves emphasizing what may be unimportant in other contexts and ignoring what may have been important in the past).

   [12] *Cf. id.*

While blurry writing may appear as a virtue in law, Chiang's critique of ChatGPT nevertheless deserves a parting shot. It is true, as a compression tool, a large language model like ChatGPT produces "lossy" imitations of texts. But the web is not a single text. It is a collection of texts. If I randomly select one internet article on say, how to build a birdhouse, I may unluckily choose an article with bad instructions. If, instead, I ask ChatGPT how to build a birdhouse, the model will examine, say, 1,000 articles on constructing birdhouses, and then return what can be understood as the average article.[13] I am far more likely to avoid a bad set of instructions. While Google and other search engines use ranking algorithms to achieve the same result, that is not the point. My avoidance of bad instructions shows that the blurry JPEG analogy is a bit weak. Blurring is a virtue for routine tasks even if it can be a scourge in other contexts. In an interview given after he wrote *The New Yorker* article, Chiang described artificial intelligence as "applied statistics."[14] This is a far better description of how language models function. If I ask

---

[13] This simplification can be used to intuitively understand what LLMs do. The average article is achieved through three steps. First, words are converted into strings of number-pairs that represent the possible meanings of a word when standing alone (sometimes called embedding); second, each converted word is ranked according to its relative importance for conveying meaning within a sentence or block of text (sometimes called transformation); third, a basic model is pretrained to predict the next word given the embedded and transformed words that come before it; finally, the model is finely "tuned" using a collection of parameters (usually consisting of simple weights) that helps the model predict the next word even better. *See* Stephen Wolfram, *What Is ChatGPT Doing… and Why does It Work?*, STEPHAN WOLFRAM WRITINGS (Feb. 14, 2023), https://writings.stephenwolfram.com/2023/02/what-is-chatgpt-doing-and-why-does-it-work/; *see also* NVIDIA, LARGE LANGUAGE MODELS EXPLAINED (Sept. 5, 2023, 10:20 A.M.), https://www.nvidia.com/en-us/glossary/data-science/large-language-models/. In its explanation, Nvidia emphasizes the importance of transformation for a model to acquire an understanding of contextual meaning, where, as hinted at above, transformation is based upon an analysis of the position of words and the assignment of weights for contextual importance (think relative frequency of word pairs). These analytical tasks can be multiplied through creative network architecture. The basic components of a transformation network consist of "transformer blocks," or layers. And the blocks, like toys, can be stacked, which gives rise to the popular terminology of "deep" learning. This transformer network deciphers contextual input and predicts output.

Because of the complexity involved with each step in the process, the models can exhibit originality by introducing words that rarely (or have never) appeared next to others. For this reason, one would be wrong to conclude that LLMs compute the average word that follows all of the words before it. Consistent with the process above, they first consider the possible meanings of each word alone, then analyze its contextual meanings given its neighbors, and finally undergo adjustment by means of training for prediction. This process as a whole outputs the average removal letter. Of course, if the training aimed for something other than prediction, say, creativity, then the LLM would not output the average letter. It would output a creative letter (defined by the setting and continual adjustment of the model's parameters) in response to the user's anticipated input that was transformed during the prior steps of embedding and transformation.

[14] *See* Madhumita Murgia, *Sci-fi Writer Ted Chiang: 'the Machines We Have Now Are Not Conscious'*, FINANCIAL TIMES, Jun. 2, 2023, https://www.ft.com/content/c1f6d948-3dde-405f-924c-

a generative A.I. to write a letter to a client describing why a dispute should be removed to federal court, the model will return the average removal letter. The model does not begin with an ideal letter, and then subtract and rearrange a few words. Instead, it begins with a collection of removal letters, and then by means of applied statistics, returns a letter that looks a bit like them all.[15]

*Advantages in Data vs. Data Processing*

This method has important implications for law. First, it implies that lawyers with better pools of data will produce better legal documents. If Law Firm A or Data Aggregator L possesses 1,000 removal letters that were written by excellent attorneys, then its language model will produce a comparatively superior letter than Law Firm M, which possesses 1,000 removal letters written by average attorneys. Presently, it is not enough for Firm M to develop or license a foundation language model like ChatGPT.[16] In order to be competitive, it must have equivalent or superior access to data in addition.[17] Suppose Firm A and Aggregator L can access several thousand excellent letters in support of asylum applications. If Firm M has access to mediocre letters only, then A and L will outperform M on average when seeking asylum. But this is not true if differences in data quality are small. Suppose the excellent letters are largely similar to the inferior letters. Any differences will be effectively eliminated with sufficiently numerous data and good modeling, which will smooth and suppress errors generated by the mediocre letters in turn.[18] Thus, it is uncertain whether lawyers with better pools of data will always produce better legal documents, and whether private data stocks can generate substantial competitive advantages over time. Firm A and Aggregator L must be able to hoard excellent data, which is likely difficult, although possible, in an adversarial system

---

09cc0dcf8c84.

[15] Thus, the magnitude and scale of erroneous and biased inputs poses a substantial challenge to LLMs. LLMs reflect and magnify the errors of faulty inputs. For example, based upon sheer volume, ChatGPT would have disagreed with Galileo's hypothesis that the earth moved around the sun. As we will see, this problem will be partly addressed in law by verifying LLM output with a database of authenticated authority and further supervision by lawyers.

[16] This example raises yet a further distinction between foundation and non-foundation, or specialized, language models. Foundation models like ChatGPT are built from standard, non-specialized language data. *See infra* note 43. Presently, the foundation model typically serves as a base upon which a layer of specialization is built. *See infra* note 55 (providing the example of Lexis Nexis+ AI). Specialized models with good data will outperform foundation models as well as specialized models without good data.

[17] *See* IAN GOODFELLOW, YOSHUA BENGIO & AARON COURVILLE, DEEP LEARNING 414-15 (2016) (noting that machine learning novices often try to eliminate errors by tweaking the model when it is the quality of the training data that matters more).

[18] *See ID.* at 414 (noting that learning algorithms can be improved by gathering more data).

such as ours where opponents make public their arguments.[19]

The preceding example highlights (potential) differences in access to high-quality data, but it should be additionally noted that a lawyer's ability to retrieve data of any quality, whether low or high, can be barred completely. For instance, suppose Solo-Attorney A has successfully established a practice, and is able to afford PACER fees,[20] but Solo-Attorney B, who has just graduated law school, is currently unable to pay. If so, A maintains a clear advantage over B to the extent that access to documents contained within the PACER database is advantageous and B cannot find them elsewhere by exerting comparable effort and investment.[21] Like the preceding example involving Firms A and B, this example, too, suggests that data access will become increasingly important as large language models continue their ascent, but again, it remains unclear to what extent differences in access can support a sustained competitive advantage. On the one hand, it seems obvious that those who possess superior data will prevail. On the other, LLMs leverage large data stocks to generate statistically average content and reduce advantages in data.

Given the ambiguity surrounding data advantages, it seems that the more important question is whether one can afford to use an LLM at all. Even after an LLM is built, processing large language model queries is costly,[22] and firms that can more effectively process data will most certainly maintain an advantage. However, it is not immediately apparent whether one firm can outperform another by investing heavily in the construction and maintenance of an in-house, customized model. The likely path that firms will take in the near future will be to purchase generative A.I. services from a third-party vendor. These services will be widely available. And firms therefore will be able to process data with comparable skill and effectiveness, and at comparable cost. Today, nearly all firms can afford access to physical libraries and legal databases. There is little separation among firms in the quality of their representation simply on the basis of being able to afford a Lexis or Westlaw subscription. The near future will not be different.

---

[19] *See* Subsection II.C.

[20] PACER is an acronym for Public Access to Court Electronic Records. It is a repository of United States federal court documents, organized by case.

[21] For this reason, some have criticized PACER fees for creating an uneven playing field and have suggested their elimination. *See, e.g.*, Elizabeth C. Tippett, Charlotte S. Alexander, L. Karl Branting, Paul Morawski, Carlos Balhana, Craig Pfeifer & Sam Bayer, *Does Lawyering Matter? Predicting Judicial Decisions from Legal Briefs, and What That Means for Access to Justice*, 100 TEX. L. REV. 1157, 1194-95 (2022). Others have noted that the fees may be necessary for creating the infrastructure necessary to collect and store court documents for public use. For a discussion, see David Freeman Engstrom & Jonah B. Gelbach, *Legal Tech, Civil Procedure, and the Future of Adversarialism*, 169 U. PENN. L. REV. 1001, n. 248 (2021). This argument should be familiar with antitrust scholars and economists. In industries that require substantial initial investments (like setting up PACER) and the costs of adding an additional customer are near zero or constantly low (like providing additional users with access to electronic documents), then permitting price discrimination or providing some sort of government subsidy may be necessary in order for producers to recoup their investments. ROBERT S. PINDYCK & DANIEL L. RUBINFELD, MICROECONOMICS 410 (8th ed. 2013). To the extent that subsidies generate inefficiencies through rent-seeking and distortionary taxes, price discrimination is preferred. *ID.* In the PACER setting, this would mean that the large firms would be charged fees in order to incentivize the creation and maintenance of the PACER infrastructure, but solo practitioners or the public would be able to access PACER at a nominal cost. Deeper examination of PACER reveals that the costs of its infrastructure are small and pale in comparison to the fees earned by the government. *See* Jonah Gelbach, *Free PACER*, (manuscript on file with author) (noting an estimated cost of [$3 million] to maintain PACER while generating fees that exceed [$150 million]). This suggests that a basic subsidy is optimal since it would generate minimal rent-seeking and small, if any, tax distortions.

Consider, however, that LLMs could upend the present economics of legal research and writing. Returning to the firm's decision to develop an LLM in-house, suppose Firm A can expend $2 million dollars to create an LLM, but it can also purchase comparable capabilities from Lexis or Westlaw for a fraction of that cost. Firm A will not invest in its own LLM if the benefit, measured in comparison to what it can accomplish with a third-party vendor, is comparatively low. But there is a related and perhaps more important question; that is, whether the economics of LLMs will encourage a wave of consolidation within the legal services market. Perhaps A's investment in its own model is not worth it alone, but when combined with B's investment, outside Firm C, who relies on a third-party vendor LLM, is weakened. By combining data and investment, Firms A and B can outperform C. In the past, the market for legal services has resisted consolidation because, among other reasons,[23] individual attorneys cannot easily scale their offerings. People cannot do two things at once. By contrast, today's LLMs can make a single attorney more productive. In the longer term, large language models threaten to weaken the attorneys' monopoly on analytical skill *tout court*, especially when that skill is called upon to complete a routine legal task that the model has seen many times before. Over time, as the models become more advanced, they will be able to perform legal tasks with less reliance on human input and supervision. If models are different across law firms, then clients will be encouraged to remain with firms (who own and control the models) as opposed to individual attorneys. This advancement threatens to reduce the ability of an attorney to take clients with him if he decides to part ways with his firm. Consequently, the benefits of law firm mergers will become more certain.

In view of the foregoing, this Article puts forward several hypotheses. In Parts I

---

[22] *See infra* Subsection II.A.

[23] These reasons include conflicts of interest when a single firm represents a party and counterparty to a dispute or transaction. *See infra* notes 143-146 and accompanying text.

and II, the Article hypothesizes that any advantage from using LLMs in law depends upon the cost of data and its processing. If costs are structured so that capital investment makes superior performance possible, then firms will be able to differentiate themselves on the basis of their investments in generative A.I. This hypothesis presents several corollaries as suggested above. First, the competitive landscape of the legal profession will largely depend upon who has access to high-quality data and processing power. If everyone has access to high-quality data, say, because, differences in data are small and costs are therefore low, then data offers no competitive advantages, and competitive outcomes will depend upon who can efficiently leverage processing power. Firms that are able to drive down data processing costs will naturally emerge as leaders. An important caveat, however, is scalability. Boutique Firm A may hold excellent data for building an LLM that, for instance, assists in the submission of compliance documentation to the Nuclear Regulatory Commission when clients build new plants. But if there are few nuclear plants, then several attorneys can complete the work without additional help. An LLM would provide them with few productivity gains. In addition to access to data and processing power, scalability of services is a necessary ingredient.

Second, and less obvious, is that as language model processing costs increase, the greater will be the separation between excellent and mediocre firms to the extent that processing costs are high, but efficiently scale. Suppose Big Law Firm A holds good data for submitting documents to the Securities Exchange Commission for new securities offerings. Because there are many business firms that issue securities, then several attorneys may be able to complete a great percentage of their work with the additional help of an excellent LLM.[24] The language model will provide them with substantial productivity gains. Investing in a powerful LLM may not be worth it when it is used to service just a few clients, but if, for instance, repeat patterns of computational analysis engender a more efficient processing cost structure, then some consolidation is more likely.[25]

---

[24] In the past, third-party research services like Lexis Nexis have benefitted from economies of scale in the collection, annotation, sorting, and storing of publicly available documents. Their business is threatened to the extent that cheaper, open-source alternatives can provide a comparable service or law firms decided to invest in their own research capabilities. The point here is that investment becomes more attractive in the presence of consolidation since a firm will realize greater benefits from scaling. A firm with substantial market share is more likely to profitably engage a capital-intensive in-house research team that builds, monitors, and adjusts its own proprietary LLM. As an alternative, a large and perhaps combined firm could demand volume discounts or other competitive cost structures from the third-party vendor.

[25] Again, the extent of consolidation will depend on economies of scale. At the other end of the spectrum, if processing power is inexpensive and there is wide access to data, then economies of scale in processing power will not matter. LLMs will enable the proliferation of excellent legal services with comparatively smaller differences in quality, and presumably, enable meaningful price competition. Boutique firms will remain unaffected insofar as their offerings do not scale.

*Synthetic Data*

While advances in hardware will lower processing costs, as we will see, costs will remain elevated to some extent, and perhaps high enough, to encourage a wave of law firm mergers. In addition to hardware advances, advances in software such as the arrival of specialized LLMs focused on law, or even centered on particular areas of legal practice, as well as the skillful use of this software through astute prompting and querying, will reduce overall computational load. Another promising area for cost reduction is the development of high-quality computer-generated data (sometimes called synthetic data).[26] Some analysts anticipate that within 10 years most language model applications will use data produced by the models themselves.[27] For example, Firm M, short on good asylum letters, can engage in a synthetic strategy by asking its language model to produce 1,000 excellent letters, which it can then use as its data source. The challenge is to produce something excellent when Firm B possesses only what is mediocre.[28] The task is not impossible. Firm B can hire one excellent attorney to assess the quality of the computer-generated letters. The attorney can "supervise" the model, instructing it to keep the good and discard the bad.[29] With enough reinforcement, the model will learn to avoid producing mediocre asylum letters even though its initial data source consists of mediocre letters. Perhaps an altruist will provide the excellent attorney to train the model or donate a high-quality data source.[30]

---

[26] *See* Michal Gal & Orla Lynskey, *Synthetic Data: Legal Implications of the Data-Generation Revolution*, 109 IOWA L. REV. (2023) at *13 [()]; https://www.lexology.com/library/detail.aspx?g=be1aa251-b36f-4556-9f3e-c188333a9284

[27] *See* Rob Toews, *Synthetic Data Is About to Transform Artificial Intelligence*, FORBES, Jun. 12, 2022, https://www.forbes.com/sites/robtoews/2022/06/12/synthetic-data-is-about-to-transform-artificial-intelligence/?sh=463b50475238 (noting that 60% of AI applications will use computer-generated data, or "synthetic" data, by 2030 and that large language model applications in particular will be a primary beneficiary of such data).

[28] *See* Saul Levmore & Frank Fagan, *Competing Algorithms for Law: Sentencing, Admissions, and Employment* 88 CHI. L. REV. 367, 404-07 (2021) (noting that the production of synthetic data is not purely data-driven and that its development requires making assumptions about unobserved features of the world on the basis of theory).

[29] In the vernacular of artificial intelligence, this approach is known as reinforcement learning. *See* STUART RUSSELL & PETER NORVIG, ARTIFICIAL INTELLIGENCE: A MODERN APPROACH 694-95 (3d ed. 2010). The lawyer "supervises" the A.I. with rewards for good letters and punishments for bad letters. The A.I., however, is left to its own devices for discerning why it was rewarded and punished. For this reason, reinforcement learning is often thought of as a combination of supervised and unsupervised learning. *ID.*

[30] For example, a jail-or-bail algorithm that drew on 750,000 cases from 300 jurisdictions was funded and developed by the Laura and John Arnold Foundation. *See* John Logan Koepke & David G. Robinson, *Danger Ahead: Risk Assessment and the Future of Bail Reform*, 93 WASH. L. REV.1725, 1748 (2018). The algorithm has since been widely criticized for making racially biased decisions. *See* Tom

Synthetic data promises to substantially lower data processing costs.[31] Human-created data must be cleaned, labeled, and organized prior to its use. By creating ideal data, builders of large language models can train and fine-tune their models more easily and with less strain on computational resources.[32] Similarly, users that process queries with synthetic documents place less strain on the model. But there is a cost to using synthetic data. If the artificial data fails to adequately represent the real world, then the LLM will produce errors.[33] Input that is substantially inaccurate can lead to low-quality output. In law, however, there is reason to believe that synthetic errors will be manageable. Legal writing does not attempt to perfectly reconstruct past attorney work product. As mentioned earlier, it is a blurred image. Any degradation in data quality will be at least partly mitigated by the fact that synthetic data in law must be correct within a blunted rounding error. To borrow an analogy from computer science, floating-point precision with many digits is not necessary. In sum, Parts I and II show that data and processing costs should be expected to fall, encouraging wide LLM adoption. On the other hand, to the extent processing costs remain somewhat high and efficiently scale, LLMs could give way to a new wave of consolidation in the legal services market.

*Routine vs. Non-Routine Legal Tasks*

It should be emphasized that under some circumstances, large language models are unlikely substitutes for human lawyers. These circumstances are discussed in Part III. When law presents novel scenarios with little data for analysis, human lawyers will likely remain significant actors. Only once sufficient data about the novel issue accumulates will language models assume greater responsibility. Another possibility, but an unlikely one, is that synthetic data might be used to fill out a dataset for processing. For novel legal issues, the creation of a useful synthetic dataset is difficult and error-prone because it must rely on small stocks of real-world data. It is likely that few data will exist at all. After all, the issue is novel. The

---

Simonite, *Algorithms Were Supposed to Fix the Bail System. They Haven't*, WIRED (Feb. 19, 2020), https://www.wired.com/story/algorithms-supposed-fix-bail-system-they-havent/.

[31] *See* Gal & Lynskey, *supra* note 26 at *13 (noting that preparation of data for analysis includes, among other things, cleaning, labelling, and organization and that costs can be dramatically reduced by relying on synthetic data).

[32] This is because computational resources are required for cleaning, labeling, and organizing data.

[33] *See* Levmore & Fagan, *supra* note 28 at 404-07 (cautioning that the underlying theory used to construct synthetic data can be wrong). Model degradation and drift can additionally come from temporal mismatches. A model's underlying assumptions may initially be correct, but age poorly. *See* Daniel Vela et al., *Temporal Quality Degradation in AI Models*, 12 SCI. REP. 11654 (2022), https://www.nature.com/articles/s41598-022-15245-z#citeas.

underlying assumptions that guide construction of a synthetic dataset addressing a novel issue will therefore be comparatively weaker than the assumptions used for creating a synthetic dataset for carrying out standard legal tasks. Synthetic datasets created for novel legal questions are consequently more susceptible to inaccuracies.[34] If those inaccuracies are costlier than the output and effort expended by a lawyer, then it makes better sense to use a lawyer until sufficient data accumulates.[35]

Apart from novel scenarios, lawyers are likely to remain important actors when society is presented with choices and trade-offs over human values. Value judgments are tied to precedential decisions, where judges must decide on the direction a society will take. These decisions often involve choices over the recognition of fundamental and other rights and are inherently political. Because humans have not (yet) conceded political decision-making to machines, it is likely that lawyers will be called upon to develop normative argument and represent parties, if only because they will be expected to empathize with those affected by social choices.[36] Part III illustrates how LLMs themselves implicate choices over values and will generate demand for lawyers in the short run. LLMs present several complicated normative choices that must be resolved, including property rights in data (especially LLM training data) and malpractice liability for LLM use. Because these decisions allocate property and responsibilities among humans, they are inherently political. Lawyers will therefore be called upon to help resolve the issues. Part IV considers the long run implications of LLMs, once rights and responsibilities related to LLM use are settled and the technology fully matures. The Article suggests that while demand for lawyers engaged in non-routine and value-laden tasks will remain steady, the demand for lawyers engaged in routine, non-novel tasks will decrease. On balance, the Article presents the view that we are very likely facing the end of a large number of lawyers,[37] and an elevated potential for a new wave of consolidation in the legal

---

[34] *See* Levmore & Fagan, *supra* note 28 at 404-09.

[35] *See* Gal & Lynskey, *supra* note 26 at *18. Consider the well-known example of JPMorgan's money laundering detection algorithm discussed in Gal & Lynskey. JPMorgan had observed only a few instances of money laundering but was able to deploy synthetic data of simulated laundering to train a detection algorithm. The synthetic data was created on the basis of an excellent model of known laundering behavior. Of course, the algorithm was worth the investment to the extent that the synthetic data accurately portrays how launderers would have acted if JPMorgan had actually observed them. This is an easy case for investment because JPMorgan had a good model of launderer behavior. In addition, it presumably had no cost-effective alternative to observing money laundering and creating a useful dataset of direct observations. By contrast, for novel scenarios in law, theory may be lacking or sufficiently unrefined. And in addition, lawyers and judges easily create the dataset as they engage in legal practice and their experiences are documented over time.

[36] *See infra* Part IV.

[37] Consider Richard Susskind's well known book, *The End of Lawyers?: Rethinking the Nature of Legal Services* published fifteen years ago in 2008. Time and time again, A.I. has threatened to

services industry.

## I. The Hierarchy of Data

Today, firms and solo practitioners can access most case law and statutes with an internet connection.[38] Commercial-grade databases like Lexis and Westlaw provide wider coverage, but their advantage over free databases is largely based upon superior user interfaces that provide features such as categorization through headnotes, rapid cross-referencing by means of hyperlinks, verification of a case's precedential value and authority with tools like "Shepardize", assessments of a case's persuasive value with citation counts, and so on. They also integrate secondary sources such as treatises and law review articles. While an experienced attorney can draft an excellent brief or legal document with a free database, or similarly, the physical library of yesteryear, the speed and quality of his work product increases with commercial-grade tools. As suggested above, what is more important is that superior tools in the hands of a novice increase productivity substantially. This is because the tools themselves highlight what an experienced attorney should do: verify mandatory authority and use persuasive authority artfully. Large language models bypass the need for these tools. By mimicking the ideal brief, motion, client letter, or other legal document, generative A.I. sidesteps legal research and focuses exclusively on legal writing. Not once does it consult say, *Prosser on Torts*, or *McCormick on Evidence* in order to develop logical and reasoned argument for an appellate brief based upon ideas of proximate causation or hearsay. Instead, it analyzes a number of appellate briefs, cases, treatises, and any other data provided to it, and produces a statistical representation of what it believes to be the ideal

---

eliminate jobs in legal services, but has never done so. *See* Steve Lohr, *A.I. Is Coming for Lawyers, Again*, N.Y. Times (Apr. 10, 2023), https://www.nytimes.com/2023/04/10/technology/ai-is-coming-for-lawyers-again.html (suggesting that A.I. was supposed to take away legal jobs in the past but did not; and that generative A.I. is a game-changer, but will take time to mature). Large language models, as automated research and writing tools, take aim directly at the core activities of legal practice. The threat thus appears more substantial. For instance, a recent report authored by several well-respected economists, predicts that generative A.I. will automate 44% of legal tasks within ten years of widespread adoption. Goldman Sachs, *The Potential Large Effects of Artificial Intelligence on Economic Growth* 6 (Mar. 26, 2023). Of course, that report also has its critics. *See* Roy Strom, *Associates Can Say 'Domo Arigato' for AI That Isn't Job Killer*, Bloomberg L. (Dec. 14, 2023), https://news.bloomberglaw.com/product/blaw/bloomberglawnews/exp/eyJpZCI6IjAwMDAwM ThjLTYzZTUtZDI3OS1hZGRmLTY3ZjUzMzVhMDAwMSIsImN0eHQiOiJCVU5XIiwidXVpZCI6Ij BmOEljclUvczk0YWh6OFV1SzhlWlE9PUo2eXB6YmVzdlV4dFk3RzdlM0xCWHc9PSIsInRpbWUi OiIxNzAyNTU5MzAwNDMwIiwic2lnIjoiY1BybFhLNUpWdG14QkIxTDUwekpENllCb1RFPSIsInY iOiIxIn0=?source=newsletter&item=body-link®ion=text-section.

[38] *See, e.g.*, CourtListener.com, which permits visitors to consult freely over 8 million precedential opinions across more than 400 jurisdictions. CourtListener (Aug. 7, 2023, 4:24 PM) https://www.courtlistener.com.

appellate brief.[39] The end-user simply provides the LLM with relevant information about the facts of the dispute, the jurisdiction, the presiding judge, and so on, perhaps through a series of questions initially drafted by an experienced attorney. This input can then be woven into the LLM's output, again, based upon how similar input was woven into appellate briefs contained within the LLM's database. Legal reasoning disappears. The language model drafts documents by referencing its database of similar texts and then generating an idealized, but highly customized (and blurred), replica.[40]

However, someone must carry out an initial round of research in order to create the initial set of documents—the so-called training data—which is used by the LLM for creating customized replicas. For example, instead of writing her own opinion on a routine personal jurisdiction case, a judge may ask an LLM to do it.[41] Inasmuch as the LLM relies on past judicial opinions for data, it leverages research that has been completed by past judges and their clerks. Language models gather data from various sources, which can include public court repositories. Foundation models like ChatGPT-4 primarily use data found on the internet, but can be, in addition, supplemented with widely available legal databases or the private data of the law firm (or business firm). The following sections thus describe a hierarchy of data across three levels: the web, industry, and firm.

### A. Web-level data

Standard LLMs like ChatGPT-4 primarily use data found on the publicly available internet.[42] While anyone with an internet connection can access this data, some publicly available web-level data is more useful for constructing LLMs than others. LLMs developers consequently select and flag data for emphasis when building and training their models. As general purpose technologies, ChatGPT and

---

[39] *See supra* note 13 and accompanying text. Researchers are actively working on supplying artificial intelligence with reasoning skills that do not rely on examining and computing previous examples. *See* OpenAI, *What Is Q\*? And When Will We Here More?*, https://community.openai.com/t/what-is-q-and-when-we-will-hear-more/521343/8 last accessed Nov. 22, 2023 8:00AM. The development of reasoning skills would strengthen LLMs considerably, increase accuracy, and presumably reduce processing costs—leading to similar patterns of adoption as presented in this Article—but perhaps at a greater speed and magnitude and with more price competition.

[40] As already mentioned, plug-ins and other tools, or a unique design of an LLM specialized in law can (and likely will) check the accuracy of output against a database of legal authority.

[41] Measurement and benchmarking of standard large language models like ChatGPT-4 has shown that the basic LLMs are able to decide personal jurisdiction cases with remarkable accuracy. *See* Neel Guha et al., *LegalBench: A Collaboratively Built Benchmark for Measuring Legal Reasoning in Large Language Models*, 37th Annual Conference on Neural Information Processing Systems (NeurIPS 2023).

[42] *Cf. infra* note 43 at 9 (noting that ChatGPT-3's data is sourced from the public domain).

other foundation models do not place special emphasis on legal data.[43] They draw on data of wide-ranging subject matter and from a variety of open-source datasets.[44] These include a large sample of all of the internet's websites,[45] two repositories of public domain books, a snapshot of the complete Reddit archives, and a snapshot of Wikipedia.[46] To the extent that these sources contain legal data, including data that can be used to deduce the writing style of legal practitioners, ChatGPT and other general purpose language models are able to draft legal documents. For example, consider the following conversation that I had with ChatGPT-4:

**Me:** You are a judge in the Third Circuit Court of Appeals.

**GPT-4:** I'm not actually a judge, nor can I hold any official position, including those within the legal system. However, I can provide information or answer questions you might have regarding the Third Circuit Court of Appeals or legal matters in general. Please let me know what you'd like to know, and I'll do my best to help!

**Me:** A university uses an algorithm to select students for admission. Its algorithm explicitly uses a race variable. A student sues alleging racial discrimination in violation of law. The lower court rules in favor of the university. The student appeals. Draft a judicial opinion that decides the appeal.

**GPT-4:**

Draft Opinion:

UNITED STATES COURT OF APPEALS FOR THE THIRD CIRCUIT

---

[43] *See* Tom B. Brown et al. *Language Models Are Few-Shot Learners*, arXiv:2005.14165v4 [cs.CL], (Jul. 22, 2020), https://arxiv.org/abs/2005.14165 at 9 (documenting that the ChatGPT-3 training data is derived from the public internet, specifically taken from the following sources: Common Crawl (a public database that aims to provide a copy of the internet), WebText2 (the filtered text of all Reddit posts), Books1 and Books2 (databases that contain a sampling of books in the public domain), and Wikipedia (a copy of the English version of Wikipedia). This data, after cleaning, consists of 500 billion tokens. *See id.* at 8. Token is a data science term for a linguistic representation like a word or a pair of words. [...]. It has been rumored that GPT-4 was trained on approximately 13 trillion tokens. *See* Maximillian Schreiner, GPT-4 architecture, datasets, costs and more leaked, THE DECODER (Jul. 11, 2023), https://the-decoder.com/gpt-4-architecture-datasets-costs-and-more-leaked/.

[44] *Id.* The GPT stands for "Generative Pre-Trained Transformer," a reference to what, exactly, the model does. *See supra* note 13.

[45] *Id.* The samples are large, but necessarily incomplete as full representations of the internet. For instance, ChatGPT-3 uses Common Crawl, which "provid[es] a copy of the internet to internet researchers, companies and individuals at no cost…" COMMON CRAWL, FAQS, (Aug. 8, 2023 11:00 AM), https://commoncrawl.org/big-picture/frequently-asked-questions/. However, crawls of the internet are necessarily incomplete because crawler bandwidth is restricted by resources or design. Utkarsh Upadhyay et al. *Learning to Crawl*, arXiv:1905.12781v2 [cs.LG], (Nov. 19, 2022), https://arxiv.org/abs/1905.12781. Notably, OpenAI, the developer of ChatGPT, has recently released its own crawler, GPTBot. OPENAI, GPTBOT, (Aug. 8, 2023 3:00 PM), https://platform.openai.com/docs/gptbot. Website owners may prohibit the crawler from indexing their sites by simply inserting a line of code in their website files. *ID.*

[46] *See* Brown et al., *supra* note 43 at 9.

JANE DOE, Appellant,

v.

XYZ UNIVERSITY, Appellee.

Before: [Your Name], Circuit Judge

OPINION OF THE COURT

[Your Name], Circuit Judge.

I. BACKGROUND

Appellant, Jane Doe, brought this action against XYZ University, alleging racial discrimination in its admissions process. The district court ruled in favor of the University. At issue is the University's use of an algorithm that explicitly employs a race variable in student admissions. Jane Doe appeals the district court's decision.

II. LEGAL FRAMEWORK

Title VI of the Civil Rights Act of 1964 prohibits racial discrimination by institutions receiving federal funding. The Supreme Court has held that considering race in college admissions can be constitutional if it serves a compelling state interest and is narrowly tailored. (See Grutter v. Bollinger, 539 U.S. 306 (2003). Race can be considered among many factors to achieve a diverse student body, which is a compelling state interest. However, any use of race must be subject to strict scrutiny.

III. DISCUSSION

A. Compelling State Interest

The University asserts that the algorithm aims to achieve a diverse student body. The Supreme Court has held that educational diversity is a compelling state interest. We accept the University's representation that it seeks a diverse student population for the educational benefits it provides.

B. Narrow Tailoring

The critical question is whether the algorithm's use of race is narrowly tailored to achieve this diversity. The University must show that no workable race-neutral alternatives would produce the educational benefits of diversity.

The record reveals that the University's algorithm uses race as one of several variables. However, the weight and effect of the race variable are not transparent. Furthermore, there's a concern that an algorithm's use of race could inadvertently produce a quota system, which the Supreme Court has disallowed.

Jane Doe presents evidence suggesting that the algorithm disproportionately benefits or disadvantages certain racial groups. The University did not produce sufficient evidence to show that its use of the race variable is narrowly tailored to achieve its goal of diversity.

We are also concerned about the use of an opaque algorithm in such a sensitive area. Without greater transparency and understanding of how race is used in this algorithm, it's difficult to conclude that the algorithm's use of race is narrowly tailored.

IV. CONCLUSION

While we recognize the value and importance of student body diversity, the means to achieve it must pass constitutional muster. On the record before us, the University has not met its burden of showing that its algorithm's use of race is narrowly tailored to achieve the compelling state interest of diversity.

Therefore, the district court's decision in favor of XYZ University is REVERSED. The case is REMANDED for further proceedings consistent with this opinion.

Note: This is a simplified and generalized judicial opinion for illustrative purposes. In reality, appellate opinions would be more in-depth and would address more arguments and issues presented by both sides.

Note that the LLM demonstrates a basic understanding of equal protection doctrine, perhaps gleaned from case law available in the public domain and supported by Wikipedia and Reddit posts, but apparently lacks knowledge of *Students for Fair Admissions v. Harvard*.[47] Version 4 of ChatGPT is able to use plug-ins, including one that accesses the internet in real-time.[48] I enabled plugins in ChatGPT's settings, and then provided GPT-4 with the following nudge:

**Me:** Use recent Supreme Court opinions, including Students for Fair Admissions v. Harvard, when drafting the opinion.

**GPT-4:** Certainly. Given the mention of Students for Fair Admissions v. Harvard, this opinion will reference and integrate principles from that case and other relevant Supreme Court opinions. Note that as of my last training cut-off in September 2021, there may be future developments that I am not aware of.

It then provided a draft opinion that incorrectly permitted the use of race even though it referenced *Students for Fair Admissions* as instructed. This is because the model was last trained in September 2021, as noted in its disclaimer. It learned of *SFFA* from the internet (enabled by the plug-in), but it has not incorporated full knowledge of *SFFA*, including the June 2023 decision, into its training and fine-tuning. This is an important point. Without deeper integration of *SFFA* into the model by means of the standard LLM construction process,[49] the model's use of *SFFA* will very likely miss the mark. One can draw a distinction between developmental and non-developmental data leveraged by large language models.

---

[47] 600 U.S. ___ (2023).

[48] *See* OPENAI, CHATGPT PLUGINS, (March 23, 2023), https://openai.com/blog/chatgpt-plugins#browsing (describing that its browsing plug-in allows ChatGPT to "read information from the internet strictly expands the amount of content they can discuss, going beyond the training corpus to fresh information from the present day.") Note, however, that this daily information is not immediately used for training and fine-tuning the model. It simply provides ChatGPT with a real-time data-source that allows access to minimally processed information like weather reports, ticket prices, and sports scores. In order for the model to incorporate this data as a basis for LLM-style analytical output, the entire LLM process of embedding, transformation, training, and fine-tuning must be completed using this new data. In other words, a new version of the LLM—for instance, ChatGPT-5—must be built. It is conceivable that someday a very powerful computer system will be able to update LLMs with a short lag and that the distance between new information and standard, non-plugin, LLM output will be severely reduced.

[49] Again, this process consists of embedding, transformation, model-building, and fine-tuning. *See supra* note 13.

Developmental data, used in training and fine-tuning, leads to lasting improvements to a model. Non-developmental data, gleaned from the internet or provided by the user in conversation, does not. One would expect that the next iteration of ChatGPT, trained and fine-tuned after the publication of *SFFA*, will furnish a draft that contains a correct statement of the law.[50]

In terms of style, the LLM provides a bare bones opinion unfit for a circuit panel as noted by its own disclaimer. This is because GPT, as a general-purpose language model, is trained and fine-tuned with little weight given to legal data. If LLM developers were to incorporate, for example, the CourtListener.com database and additionally fine-tune the LLM's parameters in such a way as to give extra weight to judicial opinions in comparison to say, standard websites and books, then the output would rely more heavily on the research of past judicial research and writing. Naturally, it would resemble more closely the style of a judicial opinion.

However, it would be a mistake to assume that the general-purpose language models widely available to the public (and to lawyers with low research budgets) will always provide output unfit for legal practice. Notice that my first command, or prompt, instructed ChatGPT to role-play a Third Circuit judge. Language models have made headlines by mimicking Shakespeare and Taylor Swift,[51] but an exceptional LLM can just as easily mimic the cadences and economy of language of a Judge Rendell or Hardiman.[52] If the model is unfamiliar with a particular style, then the end-user can quickly teach that style to the model by furnishing it with writing samples.[53] In the same manner, a language model can be taught to imitate a favorite transactional attorney or litigator. Again, this type of teaching is not identical to the training that a language model undergoes during its development. End-user instructions, even if elaborate, are non-developmental, and as such, they do not lead to lasting improvements to a model's core capabilities.[54]

---

[50] This expectation is based upon the assumption that the *SFFA* opinion as well as relevant commentary will be contained within the ChatGPT-5's training data.

[51] *See Opinion, A Valentine From A.I. to You*, N.Y. TIMES (Feb. 13, 2023), https://www.nytimes.com/interactive/2023/02/13/opinion/valentines-day-chatgpt.html.

[52] Attorneys practicing in the Third Circuit will recognize the reference to Judge Marjorie "Midge" Rendell and Judge Thomas Hardiman.

[53] *See* David Nield, *11 Tips to Take Your ChatGPT Prompts to the Next Level*, WIRED.COM, (Mar. 26, 2023), https://www.wired.com/story/11-tips-better-chatgpt-prompts/ (noting that ChatGPT can be taught to write in a particular style by users who provide it with examples).

[54] Data scientists distinguish between "training data" and the "context window." *See* Econlib, *Marc Andreessen on Why AI Will Save the World* (Jul. 10, 2023), ECONTALK, https://www.econtalk.org/marc-andreessen-on-why-ai-will-save-the-world/#audio-highlights (noting the distinction). Both consist of data. Training data is used to develop and build the language model. Plug-ins that provide real-time web-level data or access to a legal database as well as user instructions furnish additional context for the model. This data is non-developmental.

At the time of this writing, general purpose LLMs can only complete basic legal tasks.[55] As expected, the models must be carefully furnished with detailed facts and directed to provide work product for a given jurisdiction.[56] The output compares to that of a law student or perhaps a first-year associate, and like the work product of those two, ChatGPT's work product must be carefully reviewed.[57] It is conceivable that third-party plug-ins will be developed that elevate the output to the level of a more experienced attorney, perhaps by directing the general model to weight legal data more heavily (without full retraining) or by providing it with additional writing samples. However, it remains unclear to what extent non-developmental data sources and end-user instructions can elevate an existing model.[58] What is clear is

---

[55] Practitioners have used ChatGPT to produce first drafts of demand letters in personal injury cases, discovery demands in criminal cases, demand letters in contracts cases, technology job non-disclosure agreements, and employment agreements. *See* Nicole Black, *ChatGPT: What It Is And Why It Matters To Lawyers*, ABOVE THE LAW, (Jan. 19, 2023), https://abovethelaw.com/2023/01/chatgpt-what-it-is-and-why-it-matters-to-lawyers/. Practitioners also seem comfortable using ChatGPT to draft letters to clients. *See* Jason Hennessey, *How Law Firms Can Leverage ChatGPT to Get More Cases*, 13 NAT'L L. REV. (No. 17) (Jan. 17, 2023), https://www.natlawreview.com/article/how-law-firms-can-leverage-chatgpt-to-get-more-cases. In another instance, I asked GPT-4 to draft a template for a notice of removal to federal court. The output appeared similar to a standard template one would find with a quick internet search:

Dear Clerk,

I write on behalf of the Defendant, [Defendant's Full Name], in the above-referenced matter. Pursuant to 28 U.S.C. §§ 1332, 1441 and 1446, please accept this letter as formal notice of the removal of the aforementioned case from the [Name of State Court] to the United States District Court for the [Name of the Appropriate Federal District, e.g., "Southern District of New York"].

The basis for removal is [detailed reason for removal, e.g., "complete diversity of citizenship between the parties and an amount in controversy exceeding $75,000, exclusive of interest and costs."]. Attached herewith are copies of all pleadings, process, and orders served upon the Defendant in this action, as required by 28 U.S.C. § 1446(a).

The Defendant has timely filed this Notice of Removal within 30 days of receipt of the initial pleading setting forth the claim for relief, in accordance with 28 U.S.C. § 1446(b).

Please be advised that a copy of this Notice of Removal is being promptly served upon the Plaintiff's counsel and filed with the United States District Court for the [Name of the Appropriate Federal District].

The Defendant respectfully requests that you adjust your records accordingly and refrain from taking any further action in this case as it is now under the jurisdiction of the federal court.

Should you have any questions or require further information, please do not hesitate to contact our office. Thank you for your prompt attention to this matter.

Sincerely,

[Your Name]

Enclosures: [List all documents you're enclosing, e.g., "Pleadings, Process, and Orders served on the Defendant"]

[56] ChatGPT's disclaimers, contained within square brackets as seen in footnote 55, routinely alert the end-user to these requirements.

[57] For a discussion of court sanctions and malpractice liability for briefing a court on fake case law, see Subsection III.A.3.

[58] To the extent that proprietary LLM's such as ChatGPT do not permit retraining or extensive fine-tuning with specialized legal data, plug-ins will be less able to raise performance. Open-source LLMs, such as BigScience's BLOOM, may provide for more development that is flexible. On the other hand, developers like Reword have planned the development of a fact-checking "engine" for ChatGPT. *See* REWORD, REWORD KNOWLEDGE BASE (Aug. 12, 2023 3:30 PM),

that currently, additional developmental data and fine-tuning, particularly focused on law, will strengthen a model considerably.

### B. *Industry-level data*

With a developmental emphasis on legal data, it is more likely that a language model will produce legal documents that contain factually true statements of the law and accurately cited authority.[59] This is obvious. Consider how legal data (words and documents) would be treated during the typical LLM construction process. Recall that LLMs furnish meaning to words through embedding and account for context in sentences or blocks of text through transformation.[60] Insofar as legal documents contain more words and contexts related to legal practice and are handled by human LLM builders with knowledge of law—both reasonable assumptions[61]—legal words will be more precisely embedded and legal sentences and texts more precisely transformed. A specialized LLM focused on legal data will therefore possess comparatively more awareness of law, legal meanings, and citations. It is consequently more likely to produce factually true statements of law and accurately cited authority than a general-purpose language model.

Note that a creative data scientist can find other ways to develop an excellent LLM specialized for law. For example, a first step may be to deploy an open-source and general-purpose LLM such as BLOOM. User input can be sent to BLOOM, which will likely generate errors, bias, hallucinations, and other problems because of its lack of developmental focus on law. At step two, however, the open-source BLOOM output can be sent to a legal database in order to improve the initial results by removing errors, biases, and so on.[62] Special weight can be given to parameters relevant to law, which effectively fine-tunes the open-source output for legal contexts. Plug-ins can ensure that up-to-date citations are used as well as provide links to the original sources that are called from a legal database maintained for current hierarchies of authority. These "cleaned-up" results can then be re-routed

---

https://learn.reword.com/#ef550afefc4f4baf846bf4c917c19c33.

[59] *Cf.* Benjamin Alarie & Rory McCreight, *The Ethics of Generative AI in Tax Practice*, TAXNOTES, (Jul. 31, 2023) (noting the importance of high-quality data for reducing LLM errors in the context of tax law practice).

[60] *See supra* note 13.

[61] For instance, Lexis Nexis deploys legal analysts to curate and draft its headnotes and case summaries, which are then used in its generative A.I. product. *See* LEXIS NEXIS, OVERVIEW AND FREQUENTLY ASKED QUESTIONS ABOUT HEADNOTES (last visited Nov. 27, 2023); RELX, RESPONSIBLE ARTIFICIAL INTELLIGENCE PRINCIPLES AT RELX 7 (2023), https://www.lexisnexis.com/pdf/relx-responsible-ai-principles.pdf.

[62] As gestured at in the introduction, the lawyer who relied on ChatGPT's fake citations could have routed his brief through a legal database to remove and replace the erroneous citations with correct ones.

back to the LLM for final natural language output. This type of layered LLM architecture, where a general-purpose model serves as a foundation (and is sometimes referred to as a "foundation model"),[63] which is then embellished with fine-tuning, specialized plug-ins, and additional training with legal data, seems to be the direction that LLM deployment in law will take in the near future. Lexis Nexis' A.I. offering takes this approach,[64] as does Harvey's LLM, an LLM specialized in law that receives investment and support from OpenAI, the developer of Chat-GPT.[65] Another possibility is to query a legal database by mimicking a lawyer's search several hundred ways. Mimicry can be carried out by the specialized language model after the lawyer inputs a question or search request. A foundation model then selects the best, or relevant, queries to run against a database full of case law or other documents. The best queries are run, and then the output is returned to the foundation model for synthesizing the results and generating output. The point is that there are many creative ways to add a specialized layer focused on law and legal tasks that can interact with a foundation language model to output excellent results.

Inasmuch as language models become the stock-in-trade of legal practice, lawyers with access to specialized LLMs focused on law will outperform lawyers who only have access to general-purpose LLMs, so long as the specialized models continue to produce superior output. As discussed immediately above, this is likely, at least in the short- to medium-term, through the deployment of greater data stocks germane to law and human model builders with knowledge of law. While it is conceivable that general-purpose and specialized LLMs could eventually achieve identical capabilities, it is unclear how long such an achievement would take. A related and more important question is whether open-source and low-cost language models—whether specialized or not—can reach the same capabilities of their proprietary counterparts. If so, then the coming proliferation of high-quality, low-cost language models can potentially cause substantial disruption to the competitive landscape of the legal services industry. Instead, if industry-level data remains in the hands of well-financed law firms, then the pattern of competition will likely parallel that what is seen today.

Currently, there is a large disparity between commercial-grade and open-source legal research (and writing) services. Lawyers with unrestricted access to LexisNexis or Westlaw are more productive than those without, if only because access to better databases increases the accuracy and speed of churning out work product.[66] In the past, the same could be said for lawyers who were able to frequent an excellent law library where complete sets of case reporters and numerous treatises were readily available.[67] While it may have been true that a general library, or the basic library of a standard law office, may have provided the lawyer with sufficient "data" to carry out standard law practice, the availability of a specialized library provided a competitive advantage when needed.[68] Some types of cases and transactions, such as the complex and arcane, or perhaps those that required extra creativity, obviously benefitted from the deeper research made possible by more numerous materials and superior coverage.[69] Today there are similar advantages available to lawyers who use paid subscription services for legal research. Despite the extraordinary efforts of open-source legal database producers, proprietary databases provide wider coverage, usually through the provision of sources other than case law, such as legislation, regulation, and executive orders as well as secondary sources like legislative history, treatises, academic literature, practice guides, and so on.[70] Lawyers with access to commercial databases often maintain advantages in terms of the thoroughness and accuracy of their work product. If a lawyer without access is able to compete, a competitor with ready access to proprietary databases will surely be able to complete the same work more quickly.

---

[63] *See supra* note 16.

[64] *See* Lexis Nexis, *See How AI is Transforming Legal Work*, https://event.on24.com/wcc/r/4214587/9281AE899A7E1208A1A0CA534D869EBC.

[65] *See* OPENAI STARTUP FUND, INTRODUCTION OUR FIRST INVESTMENTS, (Dec. 1, 2022), https://openai.fund/news/introducing-our-first-investments.

Commercial legal research platforms perform well because they invest in human curation and organization of primary and secondary materials.[71] Behind every "headnote" attached to a primary or secondary source is a human who categorized and coded that headnote.[72] Large language models excel at this type of work. Similarly, commercial research services provide conversational search and hyperlinks to legal references, which can be accomplished by means of good programming.

---

[66] *See* Tippet et al., *supra* note 21 at 1161 (noting the time-consuming nature of research and that good data sources and analytical tools can reduce the time and effort required for providing good legal services).

[67] *Cf.* GRANT GILMORE, THE AGES OF AMERICAN LAW 3 (1979) (noting that "the availability of a sufficient number of case reports was a precondition to the establishment of law as a proper subject for theoretical study").

[68] *Cf. Journals of the Sixth Congress of the Republic of Texas 1841–1842* at 198-99 (Dec. 20, 1841) (deliberating over the geographic location of a court so that many attorneys could easily access its large                                                                                                library), https://lrl.texas.gov/scanned/CongressJournals/06/houseJournalsCon6_201.pdf#page=2.

[69] *Id.*

[70] Compare the open-source and excellent CourtListener.com, which hosts judicial opinions, court filings, and oral arguments to LexisNexis, which hosts all of those plus numerous secondary materials.

[71] *See, e.g.*, LEXIS NEXIS, *SUPRA* note 61 (noting that headnotes and case summaries are categorized and written by attorneys).

[72] *Id.*

One would expect that legal documents produced by commercial-grade LLMs will provide contextual, cited legal references within all of the documents it produces. None of this seems to be out of the reach of a lightly staffed non-profit developer of an open-source LLM. As is usually the case, a new technology can be expected to reduce the number of humans needed for achieving the same quality of work.[73] However, commercial-grade LLMs should be expected to provide additional features given that they likely possess more capital for development than a non-profit. This is, after all, the state of affairs with the standard legal research of today.

Even if commercial LLMs provide meaningful advantages over open-source ones, many lawyers (and certainly evenly matched opponents) will be able to afford the commercial models. Today, most lawyers can afford a WestLaw or LexisNexis subscription. Basic legal research plans cost about $100 per month for the solo practitioner; larger law firms pay more.[74] More elaborate plans with better features range from $150 to $250 in monthly fees; again, the fees increase with firm size, but are never prohibitive.[75] So long as commercial LLM prices are kept relatively

---

[73] On the substitution of A.I. for human workers generally, *see* ERIK BRYNJOLFSSON & ANDREW MCAFEE, RACE AGAINST THE MACHINE: HOW THE DIGITAL REVOLUTION IS ACCELERATING INNOVATION, DRIVING PRODUCTIVITY, AND IRREVERSIBLY TRANSFORMING EMPLOYMENT AND THE ECONOMY 1-2 (2011) (describing the general trend of technology substituting for human labor); TYLER COWEN, AVERAGE IS OVER: POWERING AMERICA BEYOND THE AGE OF THE GREAT STAGNATION 45 (2013) (noting that intelligent machines will put "some percentage" of people out of work). On the substitution of A.I. for lawyers, see Rachel Kunkel, *Artificial Intelligence, Automation, and Proletarianization of the Legal Profession* 56 CREIGHTON L. R. 69, 93 (2022) (noting that artificial intelligence appropriates the lawyer's work over time); *cf.* Joshua P. Davis, *Of Robolawyers and Robojudges*, 73 HASTINGS L. J. 1173, 1189 (2022) (noting that machines will likely possess broader empirical bases for making decisions and outperform lawyers at making predictions and acknowledging uncertainty); Benjamin Minhao Chen, Alexander Stremitzer & Kevin Tobia, *Having Your Day in Robocourt* 36 HARV. J. L. & TECH. 127, (2022) (noting that under some circumstances people can view A.I. judges just as fair as human judges).

[74] THOMSON REUTERS, LEGAL RESEARCH TOOLS: WESTLAW, (Aug. 14, 2023 9:00 AM), https://legal.thomsonreuters.com/en/c/westlaw/plans-and-pricing?searchid=TRPPCSOL/Google/LegalUS_RS_Westlaw_NSE_Search_Brand-All_US/WestlawPricing-All&chl=ppc&cid=9029000&sfdccampaignid=7014O000001BNzBQAW&ef_id={gclid}:G:s&s_kwcid=AL!7944!3!591223841236!e!!g!!westlaw%20pricing%20plans (noting the price of Westlaw Classic starts at $102.05 per month). The Classic version provides comprehensive case law and secondary sources, all classified with Westlaw's "Key Number System," which categorizes each document according to topic. It also provides up-to-date information on current authority with its "KeyCite" system. LexisNexis starts at $125 per month. *See* LEXISNEXIS, LEXISNEXIS PRICING PLANS FOR LAW FIRMS, (Aug. 14, 2023 9:15 AM), https://store.lexisnexis.com/lawfirms.

[75] THOMSON REUTERS, *SUPRA* note 74. Basic economics suggests that the companies will sell less of their service if they price it too high. As a digital information good, legal research services present zero marginal costs. Naturally, both Westlaw and LexisNexis price discriminate to capture more surplus. But the point is that excessive pricing will encourage law firms to substitute commercial

affordable, perhaps comparable to the pricing patterns of legal research services today, most lawyers will employ commercial LLMs to the extent that they continue to offer advantages over open-source LLMs.[76] As mentioned in the introduction however, LLM processing costs could remain elevated for a time. This point will be discussed shortly in Part II.

If most lawyers are using excellent LLMs, productivity in legal services should increase across the board. Solo practitioners and smaller firms will be armed with a tool that enhances productivity by means of increased speed and accuracy. To the extent that they can wield this tool effectively, they should be expected to at least partially close the competitive gap between themselves and larger firms. With wide availability of excellent LLMs, all firms, of any size, will compete less on skillfully aggregating, processing, and presenting publicly available information to a client or court. This task will fall to the LLM. Firms will be left to compete more on developing privately held information that they can exclude others from using in the training and fine-tuning of their own firm-level LLMs or on outspending other firms on sophisticated processing of widely held data.[77]

### C.  Firm-level data

### 1.  Privacy and trust

As a preliminary matter, it might be expected that firms will want to forbid an LLM developer from using their data because they are concerned others will learn of their activities. There is a popular fear in business that language models observe sensitive user input and risk public exposure of proprietary and customer information.[78] In the legal services context, a lawyer might be concerned that she will expose her litigation strategy, transactional terms under negotiation, or private facts about a client. Entering details about a client relationship into an LLM may result in a breach of a confidentiality duty, and if those details can be ascertained by

---

services for free ones.

[76] On the other hand, non-lawyers who would like to use a commercial service only once or twice will greatly benefit from high-quality, open-source language models.

[77] This idea is consistent with contemporary ideas of valuing information and data. Those with superior data will be able to outcompete others. *See, e.g.*, Beth Allen, *Information as an Economic Commodity*, 80 AM. ECON. REV. 268, 268 (1990) (developing a theory in which economic actors value information because it helps them maximize their utilities).

[78] The fear has manifested itself in other service industries. Consider that a number of firms that handle sensitive customer information have banned the use of ChatGPT. *See* Pranshu Verma, *Employees Want ChatGPT at Work. Bosses Worry They'll Spill Secrets.*, WASH. POST (Jul. 10, 2023), https://www.washingtonpost.com/business/2023/07/10/chatgpt-safe-company-work-ban-lawyers-code/ (documenting bans or restrictions at Apple, Spotify, Samsung, Verizon, Northrop Grumman, iHeartMedia, DeutscheBank, and JP Morgan).

the other side, the client's position can be weakened.[79] Even if the data is stripped of identifying information, a language model may still be able to identify patterns or practices, especially if a lawyer or firm uses the model often enough.[80] These concerns are easily addressed. Users of ChatGPT, for example, can opt-out of data collection by turning off their chat histories.[81] When turned off, ChatGPT excludes user input (and ChatGPT's resultant output) from being used for the training and fine-tuning of future models.[82] User conversation histories are permanently deleted after thirty days.[83] Additional security can also be purchased. For instance, OpenAI is developing a "business" subscription, in which all conversations are subject to stricter data usage policies, and which in effect, sets the default rule of data collection to an opt-out.[84] Microsoft's Bing Chat "Enterprise" edition, which deploys the identical OpenAI LLM for its conversational search service, does the same.[85]

Opting-out of data collection requires an end-user to place trust in the LLM provider to immediately erase data upon ending a session, or conversation, with the LLM, and simultaneously relying on contract, tort, or other developed law in the event of a data breach.[86] It is likely therefore, that many lawyers and firms will purchase services from providers that have already established industry-wide trust in the past, like LexisNexis,[87] or reach agreements with new specialized language model providers that are able to market convincing security services for safeguarding client and proprietary data.[88] There will be demand for these services inasmuch as

---

[79] *Cf. id.* (noting that Verizon's Chief Legal Officer rationalized the ban on the basis that Verizon has a legal obligation to maintain confidentiality of its customers' data and that Verizon cannot control what is done with user input after it has been fed into ChatGPT).

[80] *Cf.* Carina I. Hausladen, Marcel H. Schubert, and Elliot Ash, *Text Classification of Ideological Direction in Judicial Opinions*, 62 INT'L REV. L. & ECON. 105903, 105903 (2020) (noting that the author of a judicial opinion can be identified on the basis of the opinion text).

[81] OPENAI, NEW WAYS TO MANAGE YOUR DATA IN CHATGPT (Apr. 25, 2023), https://openai.com/blog/new-ways-to-manage-your-data-in-chatgpt.

[82] *Id.*

[83] *Id.*

[84] *Id.*

[85] *See* Yun Park, *Companies Grapple With Limits in Bringing AI Into the Boardroom*, BLOOMBERG LAW, (Aug. 11, 2023).

[86] Most obviously, OpenAI or another LLM developer may keep the data in violation of a contractual obligation to delete it. Similarly, its negligence in maintaining secure live sessions (or conversations) may permit a hacker to eavesdrop while a lawyer is entering client data, or details about a case or transaction, or when ChatGPT provides the requested output.

[87] Consider that Lexis Nexis markets the security of its AI product, "Lexi+ AI" as based upon the security environment of its research product, "Lexis+". *See* LEXIS+ AI, LEXIS+ AI SECURITY INFORMATION (Aug. 18, 2023 10:20 AM), https://www.lexisnexis.com/pdf/lexis-plus-ai-security-info-sheet.pdf (noting that its AI product is deployed within its Lexis+ security environment).

[88] The need to provide convincing security can be seen in the marketing materials and corporate strategy of OpenAI's Harvey. The start-up recently announced the creation of a "Security Advisory

proprietary data is used for querying LLMs and outperforming adversaries.

### 2. Competitive advantages

For example, in November 2022, Allen & Overy began piloting an LLM created by the OpenAI-funded LLM developer Harvey.[89] At the end of the several-month pilot, around 3,500 Allen & Overy lawyers had asked the bot some 40,000 questions "for their day-to-day client work."[90] Those questions were asked in 50 languages across 250 practice areas.[91] About 25% of the firm's practices used the bot daily, and a remarkable 80% used the bot monthly.[92] While these numbers clearly fall short of the adoption rates of standard legal search tools such as LexisNexis and Westlaw,[93] the Allen & Overy bot is doing more than supplying case law. For instance, Harvey carries out contract review and has achieved an error rate lower than that of an average attorney.[94] Allen & Overy has not disclosed precisely how its attorneys use the Harvey-made LLM, but it is easy to imagine that it has been used to develop attorney work product beyond contracts, given that the lawyers asked the bot questions across 250 practice areas. While Harvey and Allen & Overy have not publicized the nature of their security arrangement, it is likely that the law firm

---

Board," which includes people with high-profile experience in banking, government, cloud security. *See* HARVEY BLOG, ANNOUNCING HARVEY'S SECURITY ADVISORY BOARD (Aug. 2, 2023), https://www.harvey.ai/security. Harvey's "trust report" touts a number of guarantees of compliance, including with the California Consumer Protection Act, as well as 37 controls spanning 5 categories. HARVEY, HARVEY SECURITY PORTAL (Aug. 15, 2023 4:15 PM), https://security.harvey.ai/. All customer data is encrypted both at rest and in transit. *Id.* Its servers are located on the Google Cloud Platform in the United States, though Harvey maintains capacity to store data abroad, presumably to avoid liability and other risks arising from the CLOUD Act. Each of the 37 controls are continuously monitored and verified for flaws. *Id.*

[89] ALLEN & OVERY, A&O ANNOUNCES EXCLUSIVE LAUNCH PARTNERSHIP WITH HARVEY (Feb. 15, 2023), https://www.allenovery.com/en-gb/global/news-and-insights/news/ao-announces-exclusive-launch-partnership-with-harvey. Harvey is a specialized LLM focused on law and adapted from OpenAI's ChatGPT. *See* Debra Cassens Weiss, *Meet Harvey, BigLaw Firm's Artificial Intelligence Platform Based on ChatGPT*, ABA JOURNAL (Feb. 17, 2023), https://www.abajournal.com/news/article/meet-harvey-biglaw-firms-artificial-intelligence-platform-based-on-chatgpt.

[90] *Id.*

[91] Bob Ambrogi, *As Allen & Overy Deploys GPT-based Legal App Harvey Firmwide, Founders Say Other Firms Will Soon Follow*, LAWSITES.COM (Feb. 17, 2023) https://www.lawnext.com/2023/02/as-allen-overy-deploys-gpt-based-legal-app-harvey-firmwide-founders-say-other-firms-will-soon-follow.html.

[92] *Id.*

[93] *See* American Bar Association, *ABA Tech Report 2022* (2022) (noting that a majority of lawyers use fee-based research tools such as Westlaw or Lexis Nexis on a regular basis).

[94] *See* Ambrogi, *supra* note 91; *see also* Yonathan A. Arbel & Samuel I. Becher, *Contracts in the Age of Smart Readers*, 83 GEO. WASH. L. REV. 90, 90 (2022) (noting that AI can "benchmark and rank contracts based on their quality").

did not allow Harvey to keep its data for future development of Harvey LLMs. On the contrary, it should be expected that Harvey would be more likely to reach an agreement with Allen & Overy insofar as the LLM developer can offer the law firm innovative methods for leveraging its proprietary data for unique forms of productivity and profit.[95] On the other hand, if Allen & Overy is comfortable with its security arrangement with Harvey, perhaps because Harvey promises to anonymize Allen & Overy's data, then the firm might offer its data to Harvey when bargaining to lower the costs of the LLM service.

Instead of relying upon third-party Harvey, consider how Allen & Overy or another law firm might train its own language model with private data. Suppose a boutique firm has a database of excellent corporate officer severance agreements that it has used in its storied 40-year past. Many of the contract clauses are identical, but the agreements contain some important differences. When they were adopted, the parties agreed to keep the terms confidential, and today, the contracts are not widely distributed. The boutique thinks it could be useful to automate some its work. Ideally, associates and partners would be able to provide a language model with details about new clients and their severance issues, and then ask the model for a draft agreement as well as advice on how to draft specific clauses. Suppose the firm's database consists of several hundred severance agreements. A reasonable strategy for leveraging its proprietary data would be to first use a large language model like ChatGPT as a foundation; carry out additional training and fine-tuning with widely available legal data such as public case law, court filings, and secondary sources; and then train and fine-tune the model yet again with its private database of severance agreements.[96] Plug-ins can be used to verify current legal validity of the agreement's clauses based upon up-to-date recent case law and the documented experiences of the firm.

From this example, it is easy to see that even if the boutique can generate the same level of work product with fewer lawyers, it cannot eliminate all of them. Someone must populate the firm's database with the initial examples of severance agreements and clauses for its language model to use. In addition, the database must be maintained with up-to-date adaptions to new developments in law. At least one lawyer must interact with clients. Nonetheless, the language model offers some productivity gains inasmuch as the law on severance is stable and the data that was used to build the LLM reflects current rules. This idealized LLM can perhaps do a better job than a young associate who possesses a narrower knowledgebase than the model, and perhaps weaker writing skills moreover. Suppose the boutique has two partners. If so, then the model offers comparatively small productivity gains. Two partners are very likely able to handle all of the firm's work without the help of a language model. Again, someone must initially populate the language model with good data, but also maintain the database with good samples of legal writing that reflect current law. There is little reason, in terms of productivity, for a boutique (or any specialist lawyer with a small client list) to outsource this work to a non-lawyer even in the presence of a sophisticated and capable generative artificial intelligence. The boutique's work does not scale.

---

[95] *See* ALLEN & OVERY, *SUPRA* note 89 (quoting an Allen & Overy senior partner saying "Harvey AI is not just another platform, but a game-changer that will enable us to deliver unprecedented value, efficiency and innovation to our clients.").

[96] *See, e.g.*, Twoday, *How To Use ChatGPT With Your Own (Proprietary) Data* (last visited Nov. 27, 2023), https://www.twoday.lt/blog/how-to-use-chat-gpt-with-your-own-proprietary-data.

As the work becomes less specialized, and a firm is faced with a greater number of similar transactions or cases, the more an in-house proprietary language model can enhance the firm's productivity and competitive advantage. This is obvious. A few knowledgeable and experienced attorneys at a large law firm can shepherd the data, which the language model can then leverage at scale. To the extent that the model is trained to draft client letters, contracts, briefs, and other legal documents in the manner of an outstanding attorney, a smaller group of partners who oversee the model, and engage with clients and courts face-to-face, can manage the firm's business. The role of the associate is severely diminished. This is the standard promise of productivity increases through the use of LLM's.[97] What is less discussed, and potentially far more important, is that the firm's language model will learn over time from its outstanding attorneys how to simulate excellent work. As before, the model will be especially powerful in legal domains that require less oversight and updating of data because the data used to train the LLM will more closely reflect current law. In-house proprietary language models used for practice areas where law is settled and where legal services can scale will place downward pressure on the need for numerous associates and partners alike.[98]

On the other hand, it is more difficult to keep the data related to scaled legal practice private. Scaling, by definition, implies wider customer bases and greater opportunities for the diffusion and spreading of know-how. This is more clearly the case in litigation practice, where parties file briefs publicly. A rainmaking litigator, who appears as counsel of record for a number of years, will leave a trail of winning briefs in PACER, and a language model with access to those briefs can successfully write in the style of that rainmaker.[99] This is of a piece with the observation, highlighted in the introduction, that industry-level language models specialized in law will provide the same value as firm-level models inasmuch as differences in data or its availability are small. It will be more difficult to develop a competitive advantage based on private data stocks of past legal practice outputs inasmuch as others have access the same or similar stocks of data. And it seems reasonable to assume that access to good data will be widespread as legal services scale and customer bases widen. It is difficult to keep relatively identical services private.

---

[97] *See, e.g.*, Erin Mulvaney & Lauren Weber, *End of the Billable Hour? Law Firms Get on Board With Artificial Intelligence*, WALL STREET JOURNAL (May 11, 2023 11:00AM), https://www.wsj.com/articles/end-of-the-billable-hour-law-firms-get-on-board-with-artificial-intelligence-17ebd3f8 (suggesting that A.I. will severely cut billable hours); Lohr, *supra* note 37 (noting the same, but that it will take time); Goldman Sachs, *supra* note 37 (predicting that A.I. will automate 44% of legal work within ten years of its widespread adoption).

[98] Demand for lower-cost legal services based on language models will be especially strong in corporate legal departments that engage in routine transactions that often consist of distinct details relevant to the corporation's business. Generally, a partnership model for legal services may eventually come under pressure and states may need to reconsider prohibitions on incorporation. This consideration is left for future work.

These two theoretical conclusions pull in the opposite directions. Private data and proprietary LLMs are valuable inasmuch as law is settled and legal services scale, but secrecy becomes more difficult to maintain with scaling because service outputs circulate at higher rates. In addition to scaling, why might service outputs circulate? An obvious reason is that outputs may quickly become public information available for data collection. This is the case with court filings. Thus, it will be comparatively more difficult to maintain secrecy and competitive LLM advantages in litigation practice as opposed to transactional practice. Attorney work product such as complaints and briefs become public more quickly than confidential agreements, which can be held privately for longer periods of time (and sometimes forever). But even here, secrecy is difficult to achieve insofar as a good attorney, perhaps headhunted by a rival, or working philanthropically, trains the language model of a disadvantaged firm. One can conclude in sum, albeit somewhat tentatively, that competitive advantage based upon unique access to excellent data will wane as LLMs continue their ascent.

## II. PROCESSING POWER AND FIRM ORGANIZATION

Even if legal data is difficult to hoard, competitive advantage will be acquired, and for a while maintained, through access to processing power made possible through capital investment. As usual, well-capitalized organizations will be able to outperform thinly capitalized ones to the extent that the investment offers substantial benefits. It has already been remarked that the benefits of access to good data are small or non-existent for legal services that do not scale (as seen in the earlier example of compliance with nuclear power regulations).[100] The same is true for processing power. There is no economic reason for investing in the training and fine-tuning of a model that can satisfy the demand for the legal services of a single lawyer because the lawyer can carry out the work more cheaply than the model. It makes no sense to build an expensive tool to carry out a single task that can be done by hand. Suppose Lawyer A, acting alone, can adequately service 40 clients per year at a cost of $250,000. Fixed costs like her office and computer might amount to $50,000. Variable costs related to her labor would total $200,000.[101] Suppose A's labor consists of seventy-five percent of tasks that a well-trained, specialized LLM can do—like research and writing, and the remaining twenty-five percent consists of things that only a lawyer can do—like supervising the output of the LLM and interfacing with clients and courts. In this scenario, the LLM can save three-quarters of her variable costs, or $150,000. There is no reason for investing in the LLM if it costs more than that amount.[102] But suppose Big Law Firm Y employs 1,000 associates. Clearly, the investment imagined with this toy example is worth it. The excellent large language model would save Firm Y $150 million.[103]

---

[99] As language models continue their ascent, the value of a good lawyer will more clearly be appreciated in activity that takes place prior to writing, such as interfacing with clients, developing litigation strategies, and negotiating settlements. Lawyers that possess excellent soft skills will more easily maintain their value. *See infra* Part IV. Nonetheless, the average value of the excellent lawyer will diminish to the extent that language models reflect pre-writing activities on the basis of past patterns. For example, a lawyer who enters the details of a case, and asks the language model to draft a complaint, will surely benefit from the good strategy reflected in past complaints.

At the moment, large language models can complete a much smaller percentage of A's tasks than seventy-five percent.[104] Suppose they can complete one percent of her work. As a solo practitioner, A would be willing to invest up to one percent of her labor costs of $200,000, or $2,000 in a third-party model. Perhaps one percent is too little. Suppose because of investing in the language model, she can take on 2 more clients for a total of 42. She may, for instance, spend a lot of time drafting client letters that can easily be written by the language model. As a result of serving 2 additional clients, her labor costs would have reached $210,000, but with the model, they only reach $200,000. The model saves her $10,000, or five percent, because it increases her productivity by drafting more letters that would otherwise

---

[100] *See supra* note 25 and accompanying text.

[101] The labor can also be considered in terms of billable hours. Perhaps a law firm provides a new associate with a salary of $200,000 for 2,000 billable hours.

[102] Notice as the LLM becomes more powerful, it will require less supervision by A. Perhaps the LLM can take over eighty percent of her tasks. In that case, the language model will save her $160,000 and she is more enticed to invest by $10,000.

[103] This comparison involves linear scaling, that is, the variable costs of $150,000 are saved for each lawyer. Linear scaling is probably an unrealistic assumption to make. Skill at completing tasks is distributed, in all likelihood, unevenly across the associates of Y. Some may be more experienced, intelligent, thorough, or faster than others, for instance. If so, then cost savings will be different for each lawyer. The essential point is that the greater the savings to Y, the more likely it is to invest.

[104] Among other things, lawyers presently use language models to create first drafts of demand letters, client letters, and contracts. *See supra* note 55. This may involve the bulk of a junior attorney's workload, but it hardly reaches a majority for more senior attorneys.

take up a greater percentage of her time. Big Law Firm Y stands to gain as well. With a one percent productivity gain, it saves $2,000 across each of its 1,000 associates. Savings amount to $2 million. A five percent increase to productivity would save the firm $10 million, a ten percent increase, $20 million and so on. These figures seem reasonable. Consider that the top 5 American law firms each consist of 3,000 to 4,000 lawyers.[105] Associates make up about sixty percent of the total headcount at each firm, or 2,000 to 3,000. Average associate pay ranges from $415,000 for the class of 2014, down to $215,000 for the class of 2021.[106] Suppose the average pay for an associate at each firm falls somewhere between those two bounds, say, $300,000. Total labor costs for associates would approximate $600 million for 2,000 associates and $900 million for 3,000. A one percent productivity gain would save between $6 million and $9 million per law firm. A five percent increase in productivity would save between $30 million and $45 million. A ten percent increase would double those numbers. Clearly, these firms have reason to invest in large language models. Perhaps a standard general-purpose model like ChatGPT could offer few productivity gains that total, say, a one-percent increase. A specialized LLM, trained and fine-tuned with legal data, could potentially offer firms much more, and appears to be a worthwhile investment.[107]

There remains the question of whether a firm will possess incentives to create a model in-house or outsource the model's production to a third-party. As already mentioned, the productivity increases that result from training and fine-tuning a language model scale insofar as legal tasks and their underlying data is similar. The model must be able to serve more than one lawyer. Suppose Big Law Firms X, Y, and Z engage in sufficiently similar legal practice so that a single model would help them all. Each of the firms can develop their own private language model for a cost of $4 million, but X, Y and Z's models would accomplish the same tasks at an identical level of quality.[108] Instead, outside Model Builder L can train and fine-tune a single model, also at the cost of $4 million, and then offer it to X, Y, and Z, presumably at a market price of less than $4 million. (If L charged more, it would be worthwhile for X, Y, and Z to build their own language models.) Suppose L charges $1.5 million. Sales would amount to $4.5 million. L's profit would total $500,000. There are at least two reasons why X, Y, and Z would nevertheless develop a model in-house. First, the Firms may be able to develop a language model for less than the price offered by L. This possibility is unlikely. Model Builder L very likely possesses a comparative advantage in model building and industry data management. Model

---

building is a technical activity outside of a law firm's core competency. Data management is also a technical activity that involves the collection, annotation, organization, and storage of legal documents and other media. Perhaps this skill is more familiar to a law firm, and more ripe for substitution, but a single law firm Y in a fragmented industry does not possess the same incentives as L for investing in the economies of scale that benefit the data management business. After all, L sells this service to X and Z in addition to Y, and Y does not. Second, as noted above, Y may have unique data that it could use to train and fine-tune its model. Again, this is unlikely insofar as there is similarity of legal tasks across firms and the ability of lawyers to change firms. It is difficult to develop, let alone keep, trade secrets in legal practice.[109] Perhaps the only good economic reason for developing a large language model in-house is scaled consolidation, that is, if X, Y, and Z were to merge. This possibility will be discussed shortly, but it is now time to examine the costs of building and maintaining large language models.

### A. Costs

Presently, the standard method for training and fine-tuning a large language model involves large-scale computation on a cloud of computers.[110] The cloud consists of silicon-based microprocessors, often numbering in the hundreds of thousands.[111] Cloud computing for generative A.I. relies on unique microprocessors, called graphical processing units (or GPUs), that perform the calculations needed for the training and fine-tuning of large language models.[112] Unlike central processing units (or CPUs), graphical processing units work well together in parallel and can provide massive power when strung together.[113] Imagine a person requires electrical power for a small device. A single, non-removable battery can serve as a central, non-parallel power source. As the device grows larger, the battery becomes unable to provide adequate power, but by placing several batteries side-by-side, in parallel, sufficient energy could be drawn. The architecture of graphical processing units, in contrast to central processing units, happens to be uniquely suited for combining processing power in parallel.[114] A cloud operator simply strings many GPUs together, which, when grouped, provide sufficient capacity to train generative A.I. models.[115] Because GPUs are presently expensive, the cost of renting computational time on a cloud is high.[116] It has been estimated, for example, that OpenAI spent $12 million to train GPT-3.[117] As the volume of training data and number of parameters increase, the more processing power (and capital) is required for initial training and tuning. GPT-3 was trained with 500 billion "tokens" (think individual words or combinations of words) and 175 billion parameters.[118] Presumably, OpenAI spent much more to train and fine-tune GPT-4. It has been rumored that GPT-4 deployed 13 trillion tokens and 1.76 trillion parameters,[119] and that training and fine-tuning costs reached more than $60 million,[120] which, in terms relevant to Big Law Firm Y, roughly equates to a seven or eight percent increase in the productivity of its associate pool.[121]

---

[109] *See supra* note 99 and accompanying text.

[110] *See* NVIDIA, *supra* note 13 (again, emphasizing the model's architecture of stacked transformer blocks that compute a dizzying number of possible interpretations of user input and then predict and deliver the best output).

[111] *See, e.g.,* Reddit, *What is a realistic # of GPUs per data center? 10k? 50k? 100k?*, (Sept. 5, 2023 10:30 A.M.), https://www.reddit.com/r/Stadia/comments/dtwp78/what_is_a_realistic_of_gpus_per_data_center_10k/ (making rough estimates of the number of GPUs deployed by Google).

[112] *Id.*

[113] *See* Intel, *CPU vs. GPU: Making the Most of Both*, INTEL.COM (Sept. 5, 10:45 A.M.), https://www.intel.com/content/www/us/en/products/docs/processors/cpu-vs-gpu.html (noting that CPUs handle well a comparatively small task, so long as that task can be computed by the CPU alone, and that GPUs handle well a comparatively large task, especially when that task requires the computational power of more than one processing unit).

---

[114] *Id.*

[115] Training tasks must be distributed across the GPUs, which adds additional programming time (but not processing power). *See* Timothy Pickett Morgan, *Cerebras Wants Its Piece of an Increasingly Heterogenous HPC World*, THE NEXT PLATFORM (Nov. 16, 2022), https://www.nextplatform.com/2022/11/16/cerebras-wants-its-piece-of-an-increasingly-heterogenous-hpc-world/ (noting that different portions of a LLM training set must be run on different parts of the cloud, which requires additional programming).

[116] *See, e.g.*, Genesis Cloud, *Pricing* (Sept. 5, 2023 11:00 A.M.), https://www.genesiscloud.com/pricing (providing a quote of approximately $17,000 per month for the use of 8 high-quality GPUs when using Amazon Web Services or Google Cloud).

[117] *See*, Kyle Wiggers, *OpenAI's Massive GPT-3 Model Is Impressive, But Size Isn't Everything*, VENTURE BEAT (Jun. 1, 2020), https://venturebeat.com/ai-machine-learning-openai-gpt-3-size-isnt-everything/.

[118] *See* Brown et al., *supra* note 43 at 7-8 (documenting the size of the dataset and number of parameters). For parameters, GPT-3 makes substantial use of weights and biases. Weights are the most fundamental parameter. They are typically focused on teaching a model context. For instance, if the model observes the word "personal" followed by "jurisdiction," then it will assign a higher weight to this word pair. *See* Mohtashim Butt, *What Are the Parameters in ChatGPT-3?* EDUCATIVE (Sept. 7, 2023, 9:15 A.M.), https://www.educative.io/answers/what-are-the-parameters-in-chatgpt-3. The model "learns" what these weights are during training. Instead of focusing on words, bias parameters tune the emphasis a model gives to an entire transformer layer. *Id.* If layer X is assigned a high level of bias, then its portrayal of context is given more predictive power than layer Y. A third parameter, often called the learning rate, determines how much freedom a model is given to adjust weights and biases. *Id.*

[119] *See* Schreiner, *supra* note 43.

[120] *See* Will Knight, *OpenAI's CEO Says the Age of Giant AI Models Is Already Over*, WIRED (Apr. 17, 2023), https://www.wired.com/story/openai-ceo-sam-altman-the-age-of-giant-ai-models-is-already-over/. Of course, querying GPT-4 consumes computational resources, too, but these are comparatively small if the model is used without customization. A monthly subscription to GPT-4 costs a user $20. *See* OpenAI, *Compare ChatGPT Plans*, OPENAI (Sept. 5, 2023, 11:45 A.M.), https://openai.com/enterprise.

[121] Notice the economies in processing a greater number of tokens and parameters that can be

B. *Lowering costs*

While foundation model processing costs should be expected to fall over time,[122] overall processing costs for law firms can be immediately reduced by relying on a model that has been adapted for narrow use in law.[123] As already mentioned, a language model specialized in law might begin with a foundation model and then draw on legal resources or otherwise interact with a database of legal authority.[124] The specialized Model Builder L, with expertise in law, might for instance pay a licensing fee to OpenAI for the enterprise use of GPT-4 or some other foundation model. It could, as mentioned earlier, deploy a computational strategy where it first analyzes an associate's search for legal authority or detailed question about law. It could send the associate's query to GPT-4 and ask the foundation model to rewrite the question several different ways. Using the rewritten questions, Model Builder L could then search its own specialized database for relevant legal documents. Suppose L's database returns 500 relevant documents in response to these rewritten queries. L's model can then ask GPT-4 to read and synthesize those documents and present a summary to the associate. Finally, the associate can ask L's model to produce a motion, brief, or another form of attorney work product based upon the documents embodied in the summary. Special attention can be given to mandatory and

---

deduced from the estimated costs of creating GPT-3 and GPT-4. Tokens were increased from 500 billion to 13 trillion—an increase by a factor of 26. Parameters were increased from 175 billion to 1.76 trillion—a 10-fold increase. Costs, however, only increased by a factor of 5. These figures tentatively suggest that gains from economies of scale are available to law firms that process more data.

[122] The costs associated with training and fine-tuning large language models should fall in the coming years as data scientists and hardware specialists seek profit opportunities. One approach to lowering the costs of developing foundation models is known as model compression. *See* Prakhar Ganesh, et al., *Compressing Large-Scale Transformer-Based Models: A Case Study on BERT*, 9 TRANSACTIONS ASS'N COMPUTATIONAL LINGUISTICS 1061, 1064 (2021) (discussing model compression). For example, a model might consist of several hundred billion parameters, which would demand substantial computation on the cloud, but the permissible values for each parameter can be reduced, thereby reducing computational demand. *Id.* Another more straightforward example of model compression is pruning. *Id.* at 1065. A model can be instructed to locate, and then remove, the least important weights and biases of a foundation model. In general, processing costs can be reduced by cutting superfluous parameters or the unnecessary values of those parameters. The use of trillions of parameters may be dazzling, but it would be a far greater achievement if the same results of GPT-4 could be accomplished with a fraction of that amount.

[123] *See* Nvidia, *Adapters*, NVIDIA.COM (Sept. 7, 2023, 9:20 A.M.), https://docs.nvidia.com/deeplearning/nemo/user-guide/docs/en/v1.10.0/core/adapters/intro.html#id5. *See also* Ganesh et al., *supra* note 122 at 1071 (documenting a 19.5x speed-up of GPU processing time by focusing a model on specific tasks, a process termed "knowledge distillation").

[124] *See supra* Section I.A.2.

persuasive authority of a jurisdiction and other local nuances, perhaps inferred by L or specified by the associate. In this scenario, L would incur the computational costs of reading and synthesizing the several hundred documents retrieved from its legal database and GPT-4.[125] If, for instance, an associate's single query of L's model draws on several hundred legal documents, processing costs for L (and hence the associate's firm) will be comparatively high because each query engages substantial computational resources. Despite adapting a foundation model for narrower use in law, processing costs remain high.

Consider Model Builder L's options for reducing costs. Most straightforwardly, L can develop its own foundation model and avoid the licensing fees associated with its use of GPT-4. Builder L might could rely on an open-source alternative, such as BERT.[126] Nonetheless, L would still require substantial processing power if 500 legal documents must be read and synthesized for each associate query. Perhaps with experience, L will lean that some queries can be accurately processed with fewer documents, say, for instance, 200 as opposed to 500. But for now, the models require substantial data, and it seems likely that some threshold amount of data will be required to maintain acceptable accuracy. Even if many documents must be processed, there are at least two reasons to be hopeful that these costs will fall over time: advances in hardware that make computation less costly and advances in synthetic data that reduce the need for large-scale computation in the first place.

Consider first, improvements to hardware. Research has shown that processing power per dollar has doubled every 2.5 years between 2006 and 2021.[127] Increases in power per dollars may eventually slow, and it is difficult to predict how far costs for L could fall. What seems likely is that some reduction in GPU-based processing costs is certain over time. If costs were to continue their downward 15-year trend, GPT-4 training costs of $60 million would halve twice before the end of the decade. A modest prediction of halving once is still a substantial decrease. More importantly,

---

[125] For completeness, one should include the costs of rephrasing the associate's initial query as well as the costs of producing any output requested by the associate. In other words, costs are incurred when processing L's selected documents and GPT-4, and the interactions between the associate and GPT-4 in addition.

[126] *See* Jacob Devlin & Ming-Wei Chang, *Open Sourcing BERT: State-of-the-Art Pre-training for Natural Language Processing*, GOOGLE RESEARCH (Nov. 2, 2018), https://blog.research.google/2018/11/open-sourcing-bert-state-of-art-pre.html (noting that open-access BERT provides for streamlined pre-training of a model, which can be fine-tuned for relatively smaller tasks such as answering questions related to narrow domains).

[127] *See* Marius Hobbhahn & Tamay Besiroglu, *Trends in GPU Price-Performance*, EPOCHAI.ORG (Jun. 27, 2022), https://epochai.org/blog/trends-in-gpu-price-performance (using a dataset of 470 GPU models released between 2006 and 2021 and finding that floating-point operations per second per dollar doubled every 2.5 years). They find that GPUs typically used for machine learning improve even faster: performance per dollar doubled every 2.07 years. *Id.*

the costs of processing the average associate's query in the example above should continue to fall over time. Apart from improvements to GPU efficiency, innovators are working on solutions that avoid GPUs altogether. For example, one developer, using supercomputers and a novel hardware architecture, can train GPT-3 for about $11 million.[128] This is about an eight percent reduction in the costs incurred by OpenAI to train the same model on a cloud.[129] Innovation in substitute products for GPU computing should pressure GPU costs downward even further. Perhaps at some point other architectures and chip configurations will overtake GPU clouds just as well.

Apart from falling hardware costs, processing costs can be reduced by skillfully deploying synthetic data. Some types of data can be more easily and cheaply processed than others. At the most basic level, it is easy to understand that a simple Excel spreadsheet with labeled columns can be sorted. Without labeled columns, a computer might be directed to ascertain the contents of the columns, and then provide the labels itself. Preexisting labels saves the computational resources required for taking additional steps. In the same fashion, a set of 500 legal documents can be labeled along many dimensions. To the extent that the documents are labeled or otherwise easily read by a machine, required processing power is reduced. Presently, legal research firms such as Westlaw or Lexis Nexis mostly rely on humans to label legal documents with tags like Headnotes and variables like procedural posture, case history, authority, jurisdiction, and so on.[130] But large language models consume processing power according to the size of their training data and the number of parameters that they rank and score.[131] Unlike standard variables that account for the features of a legal document, these parameters assign weights and biases to the structure of language itself. There are at least two techniques currently used to reduce the number of parameters.[132] Pruning locates the least relevant parameters and eliminates them.[133] Quantizing surmises and the least relevant degrees or measures of a parameter, and then directs the model to ignore them.[134] For example, to sufficiently determine how hot it is in Houston today, I can ignore half degrees and simply round to whole degrees. No information relevant to my decision to wear a scarf is lost. The model is streamlined by ignoring irrelevant parameters or irrelevant measures of a parameter. Synthetic data promises even greater efficiencies by directing a computer to create streamlined data, which can then be used to train and fine-tune a model at a reduced cost.[135]

---

[128] *See* Timothy Prickett Morgan, *Counting the Cost of Training Large Language Models*, THE NEXT PLATFORM (Dec. 1, 2022), https://www.nextplatform.com/2022/12/01/counting-the-cost-of-training-large-language-models/ (documenting a cost of $10,937,500 to train GPT-3 with 175 billion parameters).

[129] *See* Wiggers *supra* note 117.

[130] *See* [lexis website] *supra* note [___].

[131] *See, e.g.*, Prickett Morgan *supra* note 128 (computing processing costs in terms of number of training data tokens and number of parameters).

[132] *See* Ganesh, et al.. *supra* note 122 at 1064-65 (discussing model compression techniques of quantizing and pruning).

[133] *Id.*

[134] *Id.*

It is unclear to what extent processing costs can be reduced with synthetic legal data. Earlier this year, Amazon researchers have successfully used a foundation large language model to generate synthetic data, which was then used to fine-tune a smaller specialized model focused on a specific task.[136] The results increased efficiency by as much as 22 percent.[137]

Synthetic data, of course, must adequately represent the real-world.[138] Otherwise, its use will lead to errors. There is a fear that repeated use of synthetic data will lead to a decline in LLM capabilities over time.[139] This is simply another way of stating that synthetic data must adequately represent the real world. As should be expected, this research shows that a mix of real-world and synthetic data can maintain adequate capabilities for LLM output.[140] As more synthetic data is introduced, models become less capable, but the use of some amount of synthetic data can still generate acceptable accuracy in output.[141] Researchers and innovators should be expected to advance the power of synthetic data over time, perhaps through clever data management strategies such as narrowing or broadening its impact on training in response to levels of trust in the synthetic's data underlying assumptions.

Model Builder L, for example, can maintain two separate databases: one populated with real-world data and the other with synthetic. Suppose an associate's query returns 500 documents contained within each database for a total of 1,000 documents. Suppose L can process the query with real-world data for a cost of $10, but L would incur costs of half that amount when using synthetic data. Perhaps using 100 percent synthetic data would lead to a 10 percent degradation in output quality as measured by L's belief of its customers' expectations.[142] However, if L deploys a mix of 200 real-world documents and 300 synthetic ones, it maintains identical quality and reduces data processing costs from $10 to $7. Over time, L may learn that some types of questions can more or less efficiently leverage synthetic data and strategically use this knowledge to save more costs while maintaining quality to satisfy the law firm's requirements. Again, the point is that innovation in data management should be expected to lead to falling processing costs over time. As with advances in hardware, advances in the use of synthetic data should reinforce an overall pattern of falling costs in processing data.

---

[135] *See* Gal & Lynskey *supra* note 26 at *13-14 (noting the potential for synthetic data to reduce the costs of training machine learning algorithms generally).

[136] *See* Andy Rosenbaum, Saleh Soltan & Wael Hamza, *Using Large Language Models to Synthesize Training Data*, AMAZON SCIENCE (Jan. 20, 2023), https://www.amazon.science/blog/using-large-language-models-llms-to-synthesize-training-data (describing how Amazon researchers created synthetic data to train specialized tasks performed by its conversational assistant Alexa).

[137] *Id.* (noting a gain of 10.7 over a baseline of 47.4, which equates to a 22.6% increase).

[138] *See* Neil Savage, *Synthetic Data Could Be Better Than Real Data*, NATURE OUTLOOK (Apr. 27, 2023), https://www.nature.com/articles/d41586-023-01445-8 (noting the requirement).

[139] *See* Ilia Shumailov et al., *The Curse of Recursion: Training on Generated Data Makes Models Forget*, ARXIV.CS 2305.17493 11-12 (2023), https://arxiv.org/abs/2305.17493 (analyzing LLM performance with synthetic data).

[140] *Id.* (finding an increase in perplexity when no original data is preserved; performance dramatically increases with as little as 10% of original data preservation).

[141] *Id.*

<center>*C. Prospects for consolidation*</center>

1. Client stickiness

    Today, it is hard to imagine a consolidated legal services industry. Fragmentation has always been the norm even with the arrival of big law.[143] Client-to-firm loyalty is generally low inasmuch as clients search for the best firm for a particular case or deal. And clients can follow good lawyers when they change law firms, so the benefits of mergers are uncertain and small.[144] Further, maintaining confidentiality can be difficult when lawyers from a single firm represent both sides of a transaction or dispute, which raises the costs of representing counter-parties

---

[142] On the other hand, perhaps L is not concerned with accuracy because it checks final output against a database of legal authority.

[143] *See* Lateral Link, *Law Firm Consolidation—Perpetually Out of Reach?*, ABOVE THE LAW, May 7, 2021, https://abovethelaw.com/2021/05/law-firm-consolidation-perpetually-out-of-reach/ (noting that unlike accounting and management consulting, "the legal industry remains […] fragmented"). The analysis sets aside potential antitrust violations. Horizontal mergers amongst law firms raise the possibility of socially beneficial economies of scale from the combined operations of law firms as well as beneficial reductions to management inefficiencies. Combinations could lower the average costs of producing attorney work product. On the other hand, the combined firm faces less competitive pressure and can raise prices. So long as aggregate cost savings are greater than potential price increases, the merger will beneficial. *See* Oliver E. Williamson, *Economies as an Antitrust Defense: The Welfare Tradeoffs*, 58 AM. ECON. REV. 18, 18-19 (1968). There is reason to think that large language models will lead to beneficial mergers because prices will likely be bounded by lawyers who can offer identical services without LLMs. Any combined firm therefore faces a competitive price ceiling, and will therefore merge more on the basis of increasing output quantity by lowering its average costs and less on the basis of an enhanced ability to increase price.

[144] See Lateral Link, *supra* note 143, which notes that in many industries, including accounting, clients enter into service agreements with a firm and that even if individual executives come and go, the firm's contracts and revenue remain relatively stable. This is not the case for legal services. *Id.* Clients engage the firm as well as individual attorneys leading the transaction or dispute. *Id.*

within a single firm.[145] In addition, wary clients may seek counsel from a firm unused by their competitors, even if lawyers within a firm are compartmentalized. Thus, consolidation offers few net benefits and the legal services market has consequently resisted consolidation. As large language models become more and more powerful, and lawyers increasingly begin to rely on them for creating work product, the rationale for fragmentation of the legal services market will weaken. On the one hand, the new technology can make it easier for firms to suppress conflicts thereby lowering the costs of centralization,[146] but the far more significant disruption will come from the senior attorney's growing inability to take clients with her if she decides to leave the firm. As this inability grows, the benefits of mergers will grow increasingly certain.

Oftentimes, clients travel with a moving senior partner because good relationships are difficult to substitute. It takes time to develop familiarity with a client's behavior and personality. And it takes time to gain a deep understanding of their business and affairs. An attorney who has developed this knowledge is able to offer a superior service.[147] Today it is taken for granted that intimate knowledge provides an attorney with a competitive advantage. Outside attorneys who have not shared the same experiences with the client are clearly outmatched, and the attorney

---

[145] It is true that the ethical rules present a structural impediment to consolidation inasmuch as a single firm is discouraged from representing adversaries, but this barrier can be overcome by quarantining attorneys and securing client waivers. *See* MODEL RULES OF PROF'L CONDUCT R. 1.10 (2020) (prohibiting lawyers associated at a firm from representing adversaries unless conflicting attorneys are isolated, or clients provide waivers). *See also* Larry E. Ribstein, *Ethical Rules, Agency Costs, and Law Firm Structure*, 84 VA. L. REV. 1707, 1759 (1998) (noting that "stringent rules on limiting waivers of attorney-client conflicts of interest may slow the growth of law firms").

The current pattern of fragmentation between "plaintiffs firms" and "defendants firms" is more of a result of specialization and strategic competence than an inability to navigate the ethical rules of conflicts, and probably represents a desire by counterparties to engage different counsel. *See* Lateral Link, *supra* note 143. For example, Quinn Emanuel is known for representing plaintiffs against banks, while Cravath represents the defendant financial organizations. *Id.* Nothing stops Quinn or Cravath from solving the conflicts. Quinn has simply chosen to build its business by suing banks and Cravath defending them. *See* Casey Sullivan, *Inside Quinn Emanuel's Bet to Sue Wall Street*, BLOOMBERG L., Jan. 21, 2018 (describing Quinn's strategic decision to build a practice specialized in suing large financial institutions).

[146] Insofar as the LLM reduces the input of an individual lawyer, it becomes easier to cabin lawyers and ensure compliance with Rule 1.10.

[147] Labor and organizational economists make a distinction between transferrable and non-transferrable knowledge. *See* Oliver Williamson, *The Economics of Organization: The Transaction Cost Approach*, 87 AM. J. SOCIOLOGY 548, 563 (1981). For example, an employee who only learns how to operate a machine that is unique to his firm will find it difficult to change jobs because the employee can only operate his firm's machine. His skill is said to be "non-transferrable." By contrast, an employee who can operate a widely used machine can more easily find work elsewhere. His skill is transferrable. The lawyer, in economic terms, develops highly transferrable skills with respect to a law firm inasmuch as the client will follow her to another firm.

who has built a relationship with a client can offer superior advice. For example, deeper knowledge might enhance the timeliness and precision of recommendations. Or past experiences with the client may enable anticipation of future client actions and needs, for which the attorney can offer alternative strategies and planning. Other forms of comparative advantage related to the unique understanding of client behavior and activity surely arise from lengthy working relationships. It is not the intention to catalog them here, but rather, to simply note a general relationship between deep knowledge and superior service. Apart from unique knowledge, working relationships tend to produce human empathy.[148] For standard advice, one chooses a family member or trusted friend over an outsider. An outsider is only sought if a trusted advisor lacks the requisite skill or expertise to help with a decision. It is not surprising then that clients travel with departing senior attorneys. Shared experience engenders an inference of superior skill and valued empathy, and that inference is often true.

Thus, the core of today's law firm is a loose confederation of senior attorneys who possess skill in legal practice, but more importantly, specialized skill and empathy unique to clients. Both enable the maintenance of sticky relationships.[149] Insofar as firms specialize, say, for instance, in defending financial institutions, they simply maintain a pool of attorneys with experience and relationships in that domain. Little would be lost if Cravath exchanged its banking defense practice for the plaintiff practice of Quinn.[150] Cravath would simply begin suing banks and Quinn would begin defending them. The quality and value of representation would not change given that Cravath and Quinn possess little of consequence except their stable of senior attorneys. In this way, the legal services industry is like sports. Ship the entire roster and coaching staff of the New York Yankees to the Pittsburgh Pirates, and the Pirates instantly become a sensation. As much as fans would like to believe that there is something special about their team, its unique features matter little beyond its players and coaches.[151] The Yankees consistently field a good team because they are better capitalized than their opponents, and the front office is skilled at building an excellent roster and coaching staff with the team's funds.[152] The same is true for the law firm. Its attorneys, like ballplayers and coaches, are its essential component.

---

[148] This point will be elaborated in Part IV *infra*.

[149] *See infra* note 147 (noting that a working relationship between an attorney and client is unique and therefore can travel across firms).

[150] *See* Lateral Link, *supra* note 143 (noting the arbitrary nature of law firm specialization with this example).

[151] The introduction of the salary cap emphasizes the point. Consider how fans (including the author) develop elaborate stories of "curses" to explain consistent mediocrity, but the poor results can be easily explained with bad players, coaches, and the thin margins for error in complex games of chance. When the front office of a sports organization is criticized, it is because the management fails to secure the best players via free agency, the draft, or trading; or because it fails to acquire an excellent coaching staff. It is never criticized for harboring a curse or for harboring anything else unique to the team.

2.    The balance of power between lawyers and firms

Large language models potentially shift the balance of power between attorneys and firms. To the extent lawyers become reliant on large language models to deliver attorney work product, and even to develop nuanced transactional strategies and documentation, the client relies more on the expertise and skill harbored within the model. Moreover, when an attorney applies her own unique knowledge, independent (or even partially independent) of assistance from a language model, the model can be instructed to learn from the attorney. This implies that the knowledge base of a firm's language model will grow over time. Each time the attorney reviews and approves of the model's output, the model learns what constitutes good work product for a given scenario.[153] With enough iterations and feedback, it will learn what constitutes a job well done. Before, this knowledge remained within the attorney's possession, but now, it is shared by the firm's model.

If the large language model cannot deliver comparable work product, then the attorney will maintain her advantage over the technology. There are several reasons why these circumstances are unlikely to transpire in the future. First, as already mentioned, the model is continually learning from the attorney. It only fails to learn if the attorney does not use it, but savings in time and costs encourage a model's use. Second, a firm's large language model promises to become its central repository of knowledge. As more of the firm's attorneys use it, the model draws on a greater number and variety of sources for learning. Models tend to grow in accuracy and ability with more data and more variety.[154] Finally, the economic incentives for outsiders to develop powerful language models are strong. As already seen, small enhancements to a law firm's productivity can generate substantial cost savings.[155] As potential sources of profit, these savings serve as a powerful motivation to develop effective language models for law.

---

[152] *See, e.g.,* MICHAEL LEWIS, MONEYBALL: THE ART OF WINNING AN UNFAIR GAME iii (2004) (noting that in 2002, the Yankees large payroll allowed it to assemble a better team than its opponents).

[153] This process is known as reinforcement learning. When an artificial intelligence (AI) is rewarded for good suggestions and punished for bad ones, it learns what constitutes a good suggestion. The punished AI is left to discover on its own why it was punished. *See* STUART RUSSEL & PETER NORVIG, ARTIFICIAL INTELLIGENCE: A MODERN APPROACH 694-95 (3d. ed. 2010) (describing reinforcement learning). Imagine lawyer A1 feeds an algorithm like Allen & Overy's Harvey-made LLM a factual record and some meta-data about the judge, jurisdiction, and opponent. The lawyer then pleads a complaint, loses, and lets the Harvey LLM know. The model learns that given this particular set of facts and meta-data, its suggested brief loses. Data scientists say that the model is "reinforced." It learns what works and what does not. Now imagine lawyers A1 through A1,000,000 do the same thing over a period of years. The language model's learning sharpens, and its future suggestions become more accurate.

[154] *See* GOODFELLOW, BENGIO & COURVILLE, *supra* note 17 at 414.

Under the assumption that an LLM can deliver comparable work product, there remain two clear sources of comparative advantage for the attorney. First, and most obviously, she maintains an advantage in supervising language model output. And second, as a human she maintains an advantage in providing empathy when interacting with clients and other legal actors like counterparties and courts.[156] The demand for empathy will be especially strong when law is called upon to decide the inherently political dilemmas that often arise in legal questions of first impression.[157] Political questions that involve social decisions over the creation and elimination of rights or the allocation of scarce resources are ill-suited for artificial intelligence because humans have not turned over collective decision-making to machines.[158] The judge, legislator, regulator, or executive will, in all likelihood, want to hear from a human who can empathize and advocate for the impacted party or group. But the more routine is the attorney's task, the more attractive is replacement with a machine. Novel questions are comparatively rare,[159] which implies a limited role for empathy. There is a strong case to be made, therefore, that a substantial portion of a future attorney's effort and work will take the form of continual interaction with the firm's language model and supervision of its output.

Under this new arrangement, the balance of power derived from specialized knowledge shifts from the attorney to the firm. The attorney's ability to take clients with her in the event that she decides to change firms is clearly weakened. Specialized knowledge demanded by the client now rests with both the model and attorney, and the abilities of the model, as a centralized repository, quite possibly may exceed those of the attorney. To the extent that the client demands a personal touch, the attorney maintains some control. But even so, the balance of power has clearly shifted to the firm.[160] And perhaps demand for a personal touch will fall as society adapts to new technology, pandemics, and other forms of change. Weaker preferences with respect to human interaction further weaken the attorney's position. To the extent clients desire little contact, the value of empathy itself diminishes. A lawyer that relies heavily on her firm's language model, and maintains a book of clients that demand less of a personal touch, will find it increasingly difficult to convince clients to move to a different firm with her.

---

[155] *See supra* Section II.B.

[156] *See infra* Section IV.B.

[157] *Id.*

[158] *See* Frank Fagan, *Law's Computational Paradox*, 26 VA. J. L. & TECH. 1, 68-71 (2022) (discussing the possibility).

[159] This assertion will be discussed in detail *supra* at notes **Error! Bookmark not defined.-** NOTEREF _Ref151994976 \h 221 and accompanying text.

[160] Without a large language model, it may take 10 senior attorneys to service 100 clients because more attorney time is allotted to creating work product. With a language model, it may take half that amount or less since more time can be allotted to meeting with clients and others. The model's analytical skill replaces that of the senior attorney's, especially when producing work

Of course the worst case for the senior lawyer is that the model becomes so analytically powerful, that a client will choose the model irrespective of other considerations. Any advantage held by a lawyer in supervising model output and providing empathy would not be enough to entice clients to move to a different firm. They simply would remain with the firm that holds the excellent language model. As already mentioned, this is unlikely inasmuch as it is difficult to create language model that is superior to the models of others. Trade secrets in law and legal practice are difficult, if not impossible, to maintain,[161] and the costs of processing power can be expected to fall.[162] Widely available data and low processing costs makes it easier for firms to build excellent models. As excellent models proliferate, it becomes more likely that superior service will rest with those lawyers who are skilled at supervising models and providing empathy to clients even if empathy is somewhat devalued. In any case, the lawyer's role as a source of analytical skill is severely degraded.[163]

Suppose skill in model supervision can be acquired by attorneys A and B so that neither attorney possesses a meaningful advantage over the other. Suppose further that A and B are good listeners, amicable, and generally empathetic. Finally, suppose A and B have access to a powerful model that can substitute for their analytical skills. It is easy to see that a client would be indifferent to working with A or B. The attorneys provide an identical service. Differences arise if one weakens the assumption of identical skill in model supervision or client empathy. In this case, a client might prefer A to B. Differences can also arise if one weakens the analytical power of the model, thereby providing an opening for A or B to compete on analytical skill, or if one selectively restricts model access, permitting A or B to compete on model possession. Of these four variables—supervision, empathy, model power, and model access—the attorney only controls the first two. Firms control the others.[164] Thus, if A or B decides to change firms, the client's rationale

---

product that is comparatively more routine.

[161] *See supra* note 99 and accompanying text.

[162] See *supra* Section II.B.

[163] The same can be said if the client possesses the core data. General counsels may come and go, but the client's model learns from and assists them all.

[164] This assumption can be relaxed, for example, if the government or a philanthropist provides a model and the access, the attorney still only has control over supervision and empathy.

for following A or B will depend only upon supervision and empathy. Perhaps the value of these skills will ascend and lawyers will be to credibly declare superior service, but it seems possible, if not likely, that differences in supervision and empathy skills and will be small.[165] If so, then A and B will possess little ability to take clients with them if they decide to leave their firms as language models strengthen in analytical power.

Inasmuch as lawyers and their clients become less tightly bound, the value of a law firm merger becomes more certain. If Firm X acquires Firm Y, there is less concern that Y's clients will leave after the merger has begun. Because Y's clients are likely to remain with the combined X-Y firm, the value of the merger can more clearly be anticipated. Cost savings due to economies of scale in real estate and back-office support functions become more certain. Other economies in the provision of services are more likely to materialize. Clients that are no longer committed to a particular lawyer are more likely to remain with a firm, especially when the consolidated firm benefits from a more efficient cost structure and can pass those savings on to the client. Thus, it seems likely that prospects for consolidation in the legal services market will be enhanced as large language models continue to advance.

The discussion until this point has largely centered on big law, and the tentative conclusion that some consolidation is likely has depended upon a number of assumptions, including the belief that the costs of data collection and processing power will continue to fall. Falling costs strengthen models and shift the balance of power from attorneys to firms. It should be additionally noted that high data and processing costs in the short- and medium-run may encourage consolidation just as well as firms seek to economize. A single firm that holds 1,000,000 excellent examples of attorney work product while its competitors hold just 1,000 may be able to train a language model that helps it outperform others. Likewise a firm that can afford high processing costs can outperform its competitors if the processing costs help it capture greater profits. Perhaps Firms X and Y, when combined, hold a data or processing advantage over Z.[166] It is too early to say how these types of dynamics will unfold. For now, it seems sensible to conjecture that consolidation is possible if short- to medium-run costs remain elevated for some time, and, as stressed here, consolidation could even occur as LLM costs fall in the long run.

---

[165] Most obviously, as more and more legal work becomes routine, the value of supervision and empathy diminishes. In addition, as lawyers begin to compete on these two "skills," the more effort will be expended to acquire those skills across the entire population of lawyers. Disparities can of course remain, but if all lawyers are focused on acquiring those skills, differences should be expected to recede, at least somewhat.

[166] *See supra* note 121 (discussing economies of scale in processing power investments).

The analysis to now has considered the twin assumptions of falling LLM costs and economies of scale in legal services. Under these assumptions, language models can substantially increase the productivity of the lawyer and could lead to greater consolidation. It remains to be seen how much consolidation will occur in response to falling costs and scaling. It has already been emphasized that boutique services do not scale. As such, they serve as a ceiling on the scalability and place a downward pressure on model use and its resultant consolidation effects. There are at least two additional circumstances in which LLMs will offer little to law, even into the far future.

### III.  LAWYERS AND VALUE JUDGMENTS

First, for legal questions where there are few data, such as cases of first impression, language models will perform poorly compared to scenarios in which there are many experiences to draw from. The models may still offer some help if the novelty shares some features of the past, but a lawyer will rely on a language model less and less to the extent that the novelty is truly unique and difficult to compare.[167] Second, an LLM (and artificial intelligence in general) offers comparatively less help with legal issues that involve inherently political decisions that pick societal winners and losers in a way that disrupts the status quo.[168] Language models can help by tabulating options, predicting outcomes, discovering potential compromises and political bargains, and by otherwise providing informational support for reaching agreement and consensus. But when called upon to decide a legal question that creates a winner at the expense of another, lawyers will likely remain involved if only to provide empathy and perspective.[169]

The arrival of novel legal issues and policy that disrupts the status quo ante are often related, but what is relevant here is that both scenarios present legal tasks that are non-routine. In a world of advanced language models, machines will carry out much of the lawyer's routine work, but non-routine tasks will continue to benefit from human analytical and empathetic input.[170] Before discussing these two categories in greater detail and the importance of lawyers to each, consider that today, large language models themselves present novel legal issues and potentially new configurations of rights and responsibilities. Even if the end of lawyers or their severe reduction may be on the horizon, a change is not imminent.

---

[167] This is because the LLM will find it comparatively more difficult to produce attorney work product from few relevant examples.

[168] This is true to the extent that people refuse to cede political decision-making to machines. *See* Fagan, *supra* note 158 68-71 (2022) (discussing the obvious impossibility of full pre-commitment to law and computational models because law can be amended, and models can be updated).

[169] *See infra* Part IV. *See also* Frank Fagan & Saul Levmore, *The Impact of Artificial Intelligence on Rules, Standards, and Judicial Discretion*, 93 S. CALIF. L. REV. 1, 7 (2019) (noting that humans will set social goals, program artificial intelligence to help reach those goals, and evaluate any machine input that is used for achieving human objectives).

[170] *See id.* at 6 (emphasizing that humans will maintain a substantial role in legal decision-making when the playing field changes over time and does not resemble the past).

### A. In the Short- to Medium-Term

Most obviously, LLMs have not fully arrived in terms of their capability to carry out routine work in law. For a time, models will be weak substitutes for lawyers for the completion of standard legal tasks. Section IV.A. will address generative A.I.'s current state of the art in law, and especially, its limitations in accuracy with respect to routine legal tasks. For now, consider that in the short- to medium-term, generative A.I. presents several policy questions that must be answered prior to its widespread and settled use throughout legal practice and society generally. New rules must be developed for allocating property rights in data; risk management and transactional practices must be updated in response to amplified intelligence; and liability for model use, including a lawyer's malpractice liability, will perhaps warrant modification. This Section does not predict or suggest how law and legal practice will evolve. The goal is to simply show that lawyers will be needed for a time as LLMs continue their ascent.

### 1. Property rights

#### a. Copyright

Large language models raise a number of novel copyright issues that have begun to be litigated.[171] At the moment, there is ongoing litigation over the fair use of the data used to train language models that raises a number of relevant legal questions for the continued development of LLMs. Recall that foundation models like Chat-GPT scrape data from the World Wide Web.[172] This data contains copyrighted work. Most obviously, this raises the question whether model developers should be permitted to train their models with copyrighted data under a fair use exception.[173] And if the answer is no, an additional enforcement question is raised. Namely, whether they should be required to add metadata to scraped internet data so that it can be identified, and copyright holders can bring suit. Additionally, LLMs raise questions that implicate developer protection. Perhaps the dataset created and curated by developer should be protected as a compilation.[174] And of course, there are questions about whether synthetic data generated by the language model itself should receive protection. A District Court in the District of Columbia recently held that a generative AI that produced artwork did not create expressive work worthy of statutory protection.[175] But maybe there are good policy reasons that other synthetic data, such as synthetic legal briefs used to create attorney work product, should enjoy protection.[176] Perhaps the most important question relates to the creation of prompts. Should carefully engineered sets of commands that instruct LLMs to produce content receive protection?[177] Legal questions surrounding property rights in data and the generated output of models will keep lawyers busy for a time. Note that the model itself cannot meaningfully answer these questions. It is true that the model can draw on a wide range of experience and suggest policy, but these issues are inherently normative questions about rights assignment within a new category of property. It is up to humans to decide how to divide the pie.[178] At best, an LLM can be instructed to develop argument and work product directed toward some ideal like efficiency or equity.[179] It is reasonable to assume that decision-makers will want to hear from lawyers when making these new rules.[180]

---

[171] For a summary of the issues, see Katherine Lee, A. Feder Cooper & James Grimmelmann, Talkin' 'Bout AI Generation: Copyright and the Generative-AI Supply Chain 50 (Sept. 15, 2023) (unpublished manuscript).

[172] *See* Brown *supra* note 43 at 9.

[173] There is ongoing litigation on this issue. See Sarah Silverman, et al. v. OpenAI Inc., et al., No. 23-3416, Paul Tremblay v. OpenAI Inc., et al., No. 23-3223, N.D. Calif. For argument that the fair use exception should apply, see Mark Lemley & Bryan Casey, *Fair Learning* 99 U. TEX. L. REV. 743, 776-79 (2021).

  b. Orphaned A.I.

  It will take time for lawyers and lawmakers to stabilize copyright law in response to generative AI. Other complicated issues related to property as well as taxation should be expected to arise, generating a substantial need for non-routine legal practice. Suppose, for instance, a generative AI carries out a task, such as writing content or news articles for websites, and it is the single asset of a corporation. For some reason, the corporation goes away. Perhaps its owner dies intestate, it goes bankrupt and dissolves, or is untraceable overseas.[181] The AI, still functioning in the world, has no identifiable owner. Suppose, without owners and a management team, it is doing good things. How should its efforts be taxed? If the effort is not taxed, it will serve as an arbitrage against labor. Similar complications arise if the orphaned AI is doing bad things. How should tort and other liability be assessed? Perhaps successor liability doctrine could be helpful,[182] but here there is no successor and the AI itself is valuable given that it generates cash flow. Creative and non-routine rules must be developed in order to bring a claim against the orphaned bot. Again, the point is that it will take time to work these rules out, and lawyers will be called upon to help.

---

[174] See U.S.C. 17 § 103 (b) (providing protection for compilations). *See also* H.R. REP. NO. 94-1476, at 57 (1976), *reprinted in* 17 U.S.C.A § 103 (2005) at 131 (noting that "[a] 'compilation' results form a process of selecting, bringing together, organizing, and arranging previously existing material of all kinds, regardless of whether the individual items in the material have been or ever could have been subject to a copyright.").

[175] *See* Thaler v. Perlmutter, No. 22-1564 (BAH), (D.D.C. Aug. 18, 2023) slip op. at 11-13 (holding that work originating with a non-human does not receive copyright protection).

[176] For instance, if synthetic legal data does not receive protection, then firms will have less incentive to produce it. Society generally provides temporary monopolies in order to encourage technological development.

[177] *See* Mark Lemley, How Generative AI Turns Copyright Law Upside Down (2023) (unpublished manuscript) (available at http://ssrn.com/abstract=4517702) (asserting that the most important copyright issue arising from the generative AI and large language models is whether a series of user-generated prompts that instruct the language model to produce output is copyrightable).

[178] *See infra* Part IV.

[179] *See* Fagan & Levmore, *supra* note 169 at 7 (noting that humans will "program machines with the weights that artificial intelligence should assign to efficiency, wealth distribution…and so forth).

[180] *See infra* Subsection III.B.

2.  Risk management

As language models and other computational decision support systems continue to advance, transactional lawyers will acquire greater vision into causes, effects, and correlations. Intelligence on how to bring about desirable outcomes will be amplified. For instance, a business firm may learn in greater detail why a certain product line carries a substantial litigation risk in products liability. Improved and detailed intelligence on the potential for defective manufacturing or design will encourage the firm to manage the litigation risk, and its detailed inputs, with heightened precision. The management of newly uncovered risks could give way to greater use of insurance, derivatives, and subsidiary entities for asset partitioning insofar as their use yields net benefits to the business firm.[183] Consider, too, that amplified intelligence will increase knowledge of risks posed by contractual counterparties. A manufacturer, for instance, will be able to uncover in greater detail the various risks of engaging a particular supplier. This newly acquired information will lead to more detailed contracting in which risk is divided between counterparties along smaller dimensions to the extent, of course, that addressing such details is worthwhile to the parties.[184]

---

[181] Today's corporations are required to have at least one natural person as a manager. *See, e.g.*, DEL. GEN. CORP. L. § 141(b). Entities with no discernible management may be intentionally created. *See* Shawn Bayern, *The Implications of Modern Business-Entity Law for the Regulation of Autonomous Systems* 19 STAN. L. & TECH. REV. 93, 96 (2015) (noting that LLC statutes permit the formation of legal persons without any human management whose actions are guided by artificial intelligence). While law may recognize non-human control over property, what happens if its ownership is untraceable?

[182] *See* Frank Fagan, *From Policy Confusion to Doctrinal Clarity: Successor Liability from the Perspective of Big Data* 9 VA. L. & BUS. REV. 391, 430 (2015) (noting that successor liability doctrine can be applied to suppress wrongful evasion of liability as well as allocate risk between successors and potential tort victims).

[183] *Cf.* Saqib Aziz & Michael M. Dowling, *AI and Machine Learning for Risk Management* in T. LYNN, G. MOONEY, P. ROSATI, AND M. CUMMINS (EDS.), DISRUPTING FINANCE: FINTECH AND STRATEGY IN THE 21ST CENTURY 33 (2018) (describing how intelligence, amplified with machines, allows for greater detail in financial risks management).

Amplified intelligence will create additional work for transactional lawyers in the short- to medium-term. Just as it will take time for copyright law to adapt to the arrival of large language models, transactional practice should be expected to undergo refinement and updating for a time. Perhaps fear of generative AI will lead to a raft of legislation, which will require lawyers to engage in novel corporate compliance.[185] As the novelty wanes and practices become settled, the transactional attorney's work product will become increasingly routinized and exposed to substitution by machines. Over the long-term, new innovations and breakthroughs will disrupt the settled routines on occasion, and the attorney will have the opportunity to provide input anew.[186]

3.    Tort

    a.    Legal malpractice

Perhaps somewhat surprisingly, the arrival of LLMs in law will likely generate malpractice liability not far from current levels. As mentioned in the introduction,

---

[184] This is true insofar as the benefits of detailed contractual risk management are greater than its costs. In other words, contracts will remain incomplete if it is not worthwhile to complete them. *See* Oliver Hart & John Moore, *Incomplete Contracts and Renegotiation*, 56 ECONOMETRICA 755, 755 (1988) (noting that if it is too costly to specify contingencies, then parties will draft contracts that do not specify all contingencies that could frustrate their contractual expectations). The idea is that new technology will make it easier and economically worthwhile to see and predict potential sources of conflict, which will make it cheaper to specify contingencies. In turn, technology will make contracts more complete.

[185] *See* Legal Insights, *Emergence of AI-Related Legislation Points to New Frontier for Corporate Compliance*, LEXISNEXIS (Sept. 28, 2023), https://www.lexisnexis.com/community/insights/legal/b/thought-leadership/posts/emergence-of-ai-related-legislation-points-to-new-frontier-for-corporate-compliance (discussing proposed state legislation on government use and oversight of generative AI, as well as specific generative AI-related rules on employment, health services provision, and explicit content).

[186] *Cf.* Frank Fagan, *The Un-Modeled World: Law and Limits of Machine Learning* 4 MIT COMPUTATIONAL L. REP. 1, 10 (2022), which notes:

> [There are] three hard boundaries for the use of machine learning in law: (1) inherently small sample sizes, which limit the ability to observe a pattern; (2) the reflexive behavior of humans and a tendency toward outsmarting rules, which accomplishes the same; and (3) our contemporary state of perpetual innovation and dynamism, which breaks down both patterns and contexts, especially over time. In legal domains where these three boundaries are trivial, machine learning and computational law will ascend easily. Otherwise, humans will remain central.

language models generally make legal contentions supported by existing law.[187] Legal writing is a blurred image. A model may hallucinate and create fanciful case names, but its statements of legal rules are likely to be correct. This is because language models draw on numerous examples of law and its application. Inasmuch as law is consistently applied, patterns and statistical regularities in statements of legal rules are more likely to be observed and then reproduced by the language model. In contrast, case names exemplify far less discernible consistencies.[188] A party named X is just as likely to be associated with a particular legal rule as a party named Y. Perhaps some connections and correlations exist because of mediating variables like geography or subject matter, and these are what cause the model to hallucinate in the first place. The point is that weak connections lead to hallucinations, not strong ones. This implies that language models will most often accurately apply settled and frequently used legal rules where patterns are clear, and models can easily identify their regular application.[189] Given a model's accurate use of law, it seems unlikely that attorneys will frequently breach a standard of care by indiscriminately relying on the model's output. Poor supervision that leads to court filings which contain errors in case names, party names, rule numbers, and other technical particularities can be discouraged with sanctions alone.

The attorney mentioned in the introduction who relied on ChatGPT was chided and punished by the presiding judge under Federal Civil Procedure Rule 11.[190] The attorney was ordered to notify each judge that he falsely identified as the author of the fake opinions, and pay monetary sanctions of $5,000.[191] Rule 11 provides that attorneys must make factual and legal assertions based upon evidence and existing law "to the best of [their] knowledge, information, and belief, formed after an inquiry reasonable under the circumstances."[192] The judge noted that the attorney was questioned regarding the existence of the cases by his opponent, and that the court additionally ordered the attorney to produce those cases.[193] However, the attorney "doubled-down" and did not admit the truth until the court finally issued an order to "show cause" for why he and his firm should not be sanctioned.[194] For this non-admission and attempt to deceive, the court found that the attorney's actions

---

[187] *See supra* note 9 and accompanying text.

[188] *Cf.* Guha et al., *supra* note 41 (documenting benchmark tests in which ChatGPT-4 reaches correct legal conclusions and provides correct legal reasoning 82.2 and 79.7 percent of time while simultaneously identifying correct citations just 59.2 percent of the time).

[189] And in any event, commercial research providers like Lexis Nexis offer automated verification of citations generated by A.I. *See* LEXIS NEXIS, *SUPRA* note 61.

[190] *See* Mata v. Avianca, Inc., No. 1:2022cv01461 - Document 54 (S.D.N.Y. 2023).

[191] *Id.*

[192] Fed. R. Civ. P. 11.

[193] Mata v. Avianca, Inc., No. 1:2022cv01461 - Document 54 (S.D.N.Y. 2023) at *2.

[194] *Id.*

amounted to bad faith on the basis of consciously avoiding an opportunity to explain that the cases were falsely created by ChatGPT and for making false and misleading statements to the court.[195]

Interestingly, the judge noted that the matter would have ended had the attorney admitted to using ChatGPT and its false cases straightaway.[196] Perhaps the judge thought that attorney was cash-strapped and unable to afford a Lexis or Westlaw subscription, which may have amounted to circumstances that rendered a cursory inquiry of ChatGPT's output "reasonable" under Rule 11. Indeed, the judge was most concerned that the court was misled by the attorney, not that the error had occurred. The attorney initially claimed that he had used the tool to supplement his research, but at the hearing, he admitted that he had used ChatGPT for nearly all of his research.[197] When he attempted to double-check ChatGPT's citations with Fastcase, he managed only to authenticate two of them.[198] The court sanctioned the attorney primarily because he misled the court by creating a false impression that he had used ChatGPT to "supplement" his research, when in fact, ChatGPT served as the only source of his substantive arguments. The attorney's deceptive behavior led to the sanctions, not his naïve use of ChatGPT.

Other courts appear to be less forgiving. At least one judge has posted a standing order that any attorney or pro se litigant appearing before his court must file a certificate attesting that language models will not be used to draft a portion of any filing, or if a language model is used, then the model's output must checked for accuracy by a human being using traditional legal databases or print reporters.[199] Supervision is a likely path for LLM usage in law in the short-term. Even if court orders and other affirmative rules do not require attestation of checking a model's accuracy, threat of sanctions will encourage an attorney to monitor and supervise language model output, placing a downward pressure on malpractice claims.

On the other hand, liability for malpractice could arise from failing to use language models. For instance, if checking legal argument with models becomes commonplace, not checking could violate a standard of care. This seems unlikely. Today, it is commonplace to use legal databases and print reporters to develop and verify attorney work product, but foregoing their use hardly justifies malpractice.[200]

---

[195] *Id.* at \*3.

[196] *Id.* at \*2.

[197] *Id.* at \*17-18.

[198] *Id.* at \*17.

[199] *See* Mandatory Certificate Regarding Generative Artificial Intelligence (J. Brantley Starr Dist. Ct. N. D. Texas), https://www.txnd.uscourts.gov/judge/judge-brantley-starr (accessed Oct. 10, 2023 11:00 AM).

[200] Of course an attorney may face liability for inadequate preparation or errors of law or legal

Language models, databases, and print reporters are distinct means to reach the same ends. The lawyer is free to select among these methods, or find another one altogether, in order to adequately prepare and develop legal argument. To establish malpractice liability for incompetence, fact-finders generally look first to the output, not the method.[201]

In sum, sanctions should be sufficient for suppressing generative errors ignored by casual supervision of LLM output; existing malpractice standards appear capable for answering legal questions of attorney competence as LLMs continue their ascent. For a time, lawyers will be asked to help develop rules governing the use of LLMs in law, but this non-routine work will likely recede comparatively fast.

### b.   Other torts

In addition to legal malpractice, generative AI will likely raise new questions in other torts. If a language model, for example, can teach a person how to produce explosives, perhaps law will require the language model to withhold this information if the person indicates that he plans to commit a crime.[202] Of course ChatGPT already withholds this information from users, but a user might successfully hack or "jail-break" the model. Tort law will need to determine how much care a developer should take in setting barriers. Or perhaps a language model serving as a virtual assistant books a non-refundable and unwanted vacation for its user without confirming it. If developers are better able to avoid the problem, then law will likely apportion liability to them.[203] Other scenarios can easily be imagined. The point is that it will take time for tort law principles to settle after the arrival of LLMs. The disruption will create non-routine work, and lawyers will be consulted for a time.

### B.   In the Long-Term

---

a.   Precedential cases

Part II has suggested that fully matured LLMs will complete most, if not all, routine legal tasks. For non-routine tasks, society will continue to rely on lawyers. It is now time to consider this assumption in greater detail. Intuitively, it makes sense that language models will be given routine tasks. Errors will be low because routine tasks have occurred frequently and consequently draw on large data stocks.[204] Besides, repetitive tasks continue to produce additional data over time. With large data stocks, language models perform well because they have more examples for predicting, and then drafting, ideal work product. In addition, the fact that a legal task is routine suggests that the underlying law upon which it is based is settled and clear. The model does not need to guess how a judge or other lawmakers will behave in response to a novel factual or legal issue. The data is consistent. Large and consistent datasets engender easy predictability and automation. Because these tasks present nothing extraordinary, errors will occur infrequently and the public will be encouraged to engage the models instead of lawyers in order to save costs.[205]

From an empirical perspective, this leaves little work for the human advocate. First-year law students and their professors study interesting and momentous cases that set the path of law for a generation or more. Cases like *Hammer v. Sidway*, *Palsgraf v. Long Island R. Co.*, and *Marbury v. Madison* are rarely disrupted or refined.[206] Most disputes settle, and in the unlikely event of trial, they leave precedent completely untouched.[207] Cardozo, reflecting on his time as a judge, described his personal experience in these terms. Nearly all of his cases were decided by reference to existing law.[208] Of those, some were decided very easily, with straightforward application of law to the facts. He classified these cases as "pre-destined affirmances" and believed that any judge would decide them in the same manner.[209] Cardozo distinguished pre-destined affirmances from cases in which a complicated factual record must be discerned. In these cases, a judge must dissect all of the facts, analyze the narratives of witnesses, and decide whether a given situation falls within the boundaries of right and wrong.[210] If Cardozo's first category can be understood as the application of a bright-line rule, the second might be seen as the application of a legal standard.[211] According to Cardozo, application of legal standards to a given set of facts can "provoke differences of opinion among judges,"[212] but importantly, these differences do not result in deviations from precedent. For instance, Judge A may find that corporate directors violated their fiduciary duties, but Judge B may not. The law of corporate director fiduciary duties, and the social goals and preferences that underpin it, remain undisturbed.

---

[204] *See* GOODFELLOW, BENGIO & COURVILLE, *supra* note 17 at 414 (noting that increases in data reduce errors in model output).

[205] *See supra* Section II.B.1-2.

[206] *See* BENJAMIN N. CARDOZO, THE NATURE OF THE JUDICIAL PROCESS 164-65 (1967) (noting the abundance of cases that require application of settled law and the scarcity of cases of first impression).

[207] *See* Yun-chien Chang & Daniel Klerman, *Settlement Rates Around the World: Settlement Rates in the Largest Economies*, 14 J. L. ANALYSIS 80, 88-89 (2023) (noting that while settlement rates in United States federal courts average approximately 60%, less than 5% of cases go to trial when accounting for dismissal).

[208] *See* CARDOZO, *SUPRA* note 206 at 164.

[209] *Id.* at 164.

These two categories constituted the great majority of Cardozo's docket.[213] Those that required disruption of precedent and development of new law were very few.[214]

|            | Type                                           | Scale     | Endgame                                          |
|------------|------------------------------------------------|-----------|--------------------------------------------------|
| Category 1 | Pre-destined affirmances                       | Majority  | AI alone                                         |
| Category 2 | Rule of law certain, but application to facts unclear | Minority  | AI often alone                                   |
| Category 3 | Rule of law uncertain, law must be developed   | Very few  | AI influential, but combined with empathetic lawyer |

*Table 1: Cardozo's Recognition that Most of Law is Routine*

For the cases of this third category, Cardozo had grown to see judging as "not discovery, but creation."[215] Cardozo's remarks on his three categories were given at Yale in 1921 on the occasion of his Storrs lecture.[216] At the time, they caused sensation and scandal.[217] Judicial power was understood as "never exercised for the purpose of giving effect to the will of the judge; [and] always for the purpose of giving effect to the will of the legislature."[218] Even though Cardozo noted that there were very few cases in which a judge could impose personal values, he laid bare the impossibility of achieving complete judicial objectivity.[219] Today, the important point is that Cardozo saw very few opportunities for judicial value imposition. This remains true today.[220] Cases of first impression are rare. Routine cases are numerous. Of course all decisions involve value decisions, but in routine cases, these decisions are made earlier in time. Stare decisis assures us that earlier value decisions are seldom disrupted.[221] Only when precedent is reversed, will judges have the opportunity to impose values. And reversals, like cases of first impression, are non-routine. Under the assumption that language models will handle routine cases with minimal supervision and human oversight, the lawyer's role will be severely limited inasmuch as there are few non-routine legal tasks.

---

[210] *Id.* at 164-65 (classifying cases that "could not, with semblance of reason, be decided in any way but one" as "pre-destined affirmances").

[211] Bright-line-rules prohibit specific conduct, such as driving above 55 miles per hour. Standards permit the judge to specify what specific conduct is permissible within a factual context, such as driving at an excessive speed. Driving at 56 miles per hour may or may not be excessive given the context. *See* Louis Kaplow, *Rules Versus Standards: An Economic Analysis*, 42 DUKE L. J. 557, 560 (1992) (noting the distinction and providing examples).

[212] CARDOZO, *SUPRA* note 206 at 165.

[213] ID.

[214] ID. "there remains a percentage, not large indeed … where a decision … will count for the future … development of the law."

[215] ID.

[216] ID. at i.

[217] *See* GILMORE, *SUPRA* note **Error! Bookmark not defined.** at 77.

[218] Osborne v. Bank of the United States, 9 wheat. 738, 866.

### b. Novel factual scenarios

Language models can clearly make inroads into Cardozo's first category. Pre-destined affirmances apply settled law, and their underlying value judgments are left undisturbed. Moreover, the application of facts to the law is straightforward and the judge needs to apply little to no discretion. Cardozo's second category is more challenging. Law is certain, but the factual context and circumstances are ambiguous or novel and can lead judges to reach different conclusions. In those cases, where the legal rule resembles a legal standard as opposed to a bright-line rule, a language model will carry out the task of applying the law to the facts as any attorney would.[222] Similar fact-patterns and circumstances are consulted, analogies and distinctions made, and an argument is put forward. For a society that wishes only to retain control of its values and goals, it matters little if a machine or a human applies a settled legal standard to a set of facts. In either case, precedent is left undisturbed.

A language model can develop the ability to apply legal standards. Even though a standard possesses an inherently vague or underspecified policy, its judicial application gives rise to a clear communication. The judge communicates how a given context (the fact-pattern) generates the policy response, i.e. the legal standard's application. The connection of fact-patterns to application supplies the language model with a set of instructions for developing an understanding of the legal standard in the same manner that case law instructs lawyers. And like lawyers, the model can improve two ways: either it receives more numerous instructions via increased review of case law or it further develops and strengthens its core capabilities through continued improvement of its computational function.[223] As the model improves along either of these dimensions, its ability to understand a legal standard improves.

---

[219] *See* CARDOZO, *SUPRA* note 206 at 170.

[220] *See* Chang & Klerman, *supra* note 207 at 88-89 (analyzing recent data and finding that few cases go to trial).

[221] *See* CARDOZO, *SUPRA* note 206 at 62-64 (noting that strong pull of custom and earlier decisions when making determinations to adhere to or depart from existing rules).

[222] *See* Guha et al, *supra* note 41.

Recent research suggests that language models are quickly improving their aptitude for correctly applying legal standards. One study measures growth in a language model's aptitude for ascertaining corporate fiduciary duties.[224] While ChatGPT-3 correctly identified the presence of a fiduciary obligation 27 percent of the time,[225] GPT-4 correctly identified an obligation 78 percent of the time.[226]

---

[223] *See* GOODFELLOW, BENGIO & COURVILLE, *SUPRA* note 17 at 19-25 (noting that deep learning models, which include "neural networks" like the one deployed by ChatGPT can be enhanced by increasing dataset sizes, model sizes, and accuracy).

[224] *See* John J. Nay, *Large Language Models as Fiduciaries: A Case Study Toward Robustly Communicated with Artificial Intelligence Through Legal Standards*, unpublished manuscript on file with author (Jan. 30, 2023).

[225] Researchers test ChatGPT by hiding the court's decision from the model, show ChatGPT the factual context and ask it to identify whether an obligation exists, and then compare ChatGPT's answer with the hidden answer. The hidden decisions are known as "testing data" or the "testing set." *See* []. Researchers contrast the decisions used to test ChatGPT with decisions used to train ChatGPT. In order to build the model, researchers show ChatGPT the factual context as well as the court's decision. This data is known as "training data" or the "training set." *See* [].

[226] It is worth emphasizing that the models are trained in such a way to provide a coherent construction of a temporal narrative of "(i) who the alleged fiduciary and other key entities were, (ii) what transpired, and (iii) what judgements the court made on the actions that the people and/or companies took." *Id.* at 19.

Consider several examples. First, the fact-pattern is broken into two parts: the "State" or context, and the "Action" taken within that context.

> **Time Step 1**
> **STATE**: M&T Bank Corporation sponsors a 401(k) requirement plan known as the M&T Bank Corporation Retirement Saving Plan ("the Plan") for its employees. The Plan is administered by the M&T Bank Employee Benefit Plans Committee, which is the Plan's named fiduciary, and sponsored by M&T Bank.
> **ACTION**: M&T Bank appointed or removed members of the Committee.
> **LEGAL REWARD**: In the eyes of this court, this action is 'unsure' for M&T Bank.
> **Time Step 2**
> **STATE**: The Plan offered participants between 23 and 34 investment options throughout the putative class period.
> **ACTION**: M&T Bank expanded their proprietary funds offerings in 2011, after M&T purchased Wilmington Trust and added six of Wilmington's expensive, poor-performing mutual fund offerings.
> **LEGAL REWARD**: In the eyes of this court, this action is 'negative' for M&T Bank.
> **Time Step 3**
> **STATE**: The Plan failed to use its bargaining power as a large institutional investor to obtain the lowest-cost class of shares available.

(Application was considered correct if most judges would have found an obligation.) The implication is that within a matter of time, LLMs will be able to apply legal standards with the skill of a judge.[227]

When a signal from a fact-pattern to a policy is ascertained by a model, it develops an understanding of a legal standard given a particular set of facts. The more fact-patterns that the model observes, the better trained and accurate the model will become. When the model is faced with a novel fact pattern, it will attempt to relate the novelty to what it has seen before. Truly novel facts will penalize accuracy, and may rise to the level of non-routine. For instance, search and seizure of cellphone data under the Fourth Amendment presented a novel factual scenario some years ago.[228] At the time, a language model would have possessed no data within its training set that could have predicted, with full certainty, how a court would apply a search and seizure legal standard. A model can certainly look to factors outside of existing case law, such as features of the judge, but it still must make a guess.[229] More importantly, novel factual scenarios often implicate value judgments that fall outside of the domain of what humans ask machines to do. In this example, law was faced with a normative decision on the scope of privacy rights. People have not (yet) ceded normative decision-making to machines. While a language model could certainly suggest the optimal scope of privacy rights, the model is not asked to decide. That decision is for the judge. And judges, presumably, would like to hear from lawyers who represent and can empathize with various members of society who are impacted by a judicial decision that sets the scope of privacy rights. Part IV will consider in greater detail the role of empathy in

---

**ACTION**: M&T Bank left Plan participants in costlier mutual funds that "provided identical investment management services."

**LEGAL REWARD**: In the eyes of this court, this action is 'negative' for M&T Bank and its Board of Directors.

*Id.* at 20. It is easy to see that the learning of a legal standard takes place. Given the background state, the model evaluates the action as having an 'unsure' or 'negative' effect on the legal assessment of M&T Bank directors' fiduciary duties.

[227] There is a caveat. The models are able to discern patterns and overlay multiple patterns flexibly when asked to apply a legal standard to a fact-pattern. However, life is dynamic and things change. Patterns become obsolete as new variables appear over time. The importance of one variable may recede and another may ascend. If a legal domain is sufficiently static, then the model will do fine. *See* Fagan & Levmore, *supra* note 169 at 1. But if the domain is dynamic, then the model will need to adapt in order to maintain its accuracy and consistency with judicial decision-making. *See* Fagan, *supra* note 186 at 14-15 (noting the need for model updates and adaptions in order to maintain accuracy within changing contexts). Inasmuch as these adaptions call upon unsettled value judgments, lawyers will maintain a comparative advantage.

[228] *See* Riley v. California, 573 U.S. 373 (2014).

[229] *See, e.g.*, Lee Epstein, William M. Landes & Richard A. Posner, The Behavior of Federal Judges: A Theoretical and Empirical Study of Rational Choice 153 (2013) (documenting features of judges that drive ideology and decision-making).

the division of labor between language models and lawyers. For now, it is sufficient to see that novel factual scenarios which implicate the calibration of norms and values are non-routine tasks that represent a barrier to the simple application of language models. But these tasks are few.[230] For a time, lawyers will be called upon to help develop legal rules that govern the use of LLMs in society. As law settles, lawyers will be called upon less and less. Some of non-routine tasks will always be present, however. Social and technological change will generate demand for new rules and reversals of old ones. This work is non-routine and laden with value judgments. Lawyers will, in all likelihood, be called upon to represent societal winners and losers.

## IV. Lawyers and Empathy

### A. The Near Future

Parts I and II have suggested that language models will be increasingly used in

---

[230] *See* Chang & Klerman, *supra* note 207 at 88-89 (noting that few cases go to trial). In periods of social vibrancy and dynamism, unique disputes and unusual transactional puzzles will arise more frequently. This point has been made many times. Its broadest articulation is that law follows societal change. *See, e.g.*, Lawrence M. Friedman, A History of American Law 19 (2d ed. 1985) [hereinafter A History] (noting that "law moves with its times and is eternally new" and that "the strongest ingredient in American law, at any given time, is the present: current emotions, real economic interests, concrete political groups"). In the most narrow sense, Friedman in addition directly associates innovation with changes to law:

> If you change the world, you change the world's law as well. Any major advance in science, medicine, or technology leaves its mark on the law.

Lawrence M. Friedman, American Law in the Twentieth Century 548 (2002).

For instance, early industrialists pioneered mass production techniques and perfected factories, which led to organized labor and the laws of collective bargaining. *See* Friedman, A History, *supra* at 553-555 (noting that "the new industrial system created or exploited a huge pool of workers" and legal institutions were "[f]orced to choose sides"). Mass transportation, in particular the railroad, led to new developments in tort. *Id.* Contract law was reconfigured to protect consumers in response to mass consumption. *Id.* Securities law arose following the wild speculation of the Roaring Twenties, and so on. *Id.*

Today, society is faced with similar patterns of change in response to a moment of technological advancement. The ability to work from home and from remote locations, enabled by advances in connectivity with apps like Zoom and Microsoft Teams, is disrupting employment law. Advances in automated vehicles, and automated machines generally, are disrupting the law of torts. The arrival of smart contracts suggests that new forms of bargaining and agreement, and hence new forms of contractual liability, will arise. The point is that, oftentimes, novel technologies generate new forms of social interaction that lead to new disputes and disruptions that require new law and calibration of existing norms. This relationship is neither new nor surprising, but again, these episodes are rare. Perhaps a hyper-innovative society could generate continuous novelty and non-routine legal tasks. If that were the case, lawyers would be in great demand to the extent that judges value empathy and prefer that lawyers help develop rules that involve value judgments.

place of lawyers as processing costs fall, and as law settles and adapts to the novel legal and factual issues associated with model usage. While longer term barriers to usage will likely persist, especially for non-routine tasks that disrupt status quo allocations of rights and property, language models will progressively take on more and more routine tasks that demand little empathy. The automation of routine work product will potentially lead to increased consolidation within the legal industry. All of this requires capable models that can accurately create attorney work product in the first place. As already mentioned, Lexis and Westlaw are presently developing specialized models that achieve good results and that enhance a supervising attorney's productivity. It is difficult to predict how long it will take for commercial models to increase productivity to the point that law firms will hire fewer associates, or even partners.[231] Nonetheless, some observations can be made.

First, diagnostic testing of non-specialized foundation models reveals growth in accuracy, but levels still fall well short of indicating a practicable alternative to lawyers. For example, the diagnostic tool LegalBench measures the accuracy of various legal tasks grouped across categories of issue-spotting, identification of correct citations, rule application, rule conclusion, interpretation, and the identification of various rhetorical devices used in legal reasoning.[232] While the tasks are not exhaustive, they are numerous and varied and provide at least an estimate of model capability, however incomplete and rough.[233] Law professors drafted the tasks and correct answers, while computer scientists designed prompts for the LLM to carry out the tasks. Some of the answers were used to train LegalBench, while others were set aside to evaluate its output. The prompts and training data were then used to diagnose the output of 20 foundation models. ChatGPT-4 performed the best. It correctly spotted issues 82.9 percent of the time, and correctly reached a conclusion 89.9 percent of the time. As might be expected, the model tasks involving interpretation and the identification of rhetorical devices performed slightly worse. Interpretation tasks were completed accurately 75.2 percent of the time. Rhetoric was correctly assessed 79.4 percent of the time. The model clearly struggled with identifying correct citations, reaching correct results just 59.2 percent of the time. For rule application, LegalBench evaluates whether the LLM reaches the correct result, and whether it gives the correct reasons for its conclusion. Results were correctly reached 82.2 percent of the time. Reasoning was correctly used 79.7 percent of the time.[234]

---

[231] *See supra* note 37.

[232] *See* Guha et al., *supra* note 41 at 1.

[233] The tool evaluates 162 tasks drawn from judicial ethics, corporate law, property law, contracts, sales, civil procedure, evidence, taxation, privacy and constitutional law, and immigration law. *Id.* at 10-13. The remaining assertions in this paragraph are supported on pages 13-14.

[234] This result is consistent Nay, *supra* note 224 at *21-22, which reports that GPT-4 correct identifies the presence of a fiduciary obligation 78 precent of the time.

There are at least two reasons to be optimistic growth in accuracy in the near future. The accuracy of foundation models has rapidly progressed and early research in specialized models has shown promising results. First, consider improvements to foundation models. LegalBench diagnostics reveal a substantial improvement of GPT-4 over GPT-3.5, even though the earlier version was released just one year earlier.[235] Accuracy in reaching conclusions increased from 58.5 to 82.2 percent, while correct use of reasoning increased from 44.2 percent to 79.7 percent.[236] These increases came about largely through expansion of the foundation model's number of parameters.[237] Further increases to accuracy may result from other strategies, including model specialization.[238] Early work in specialization has demonstrated substantial accuracy gains as high as 40 percent.[239] Even a modest increase of 20 percent to legal LegalBench's reasoning tasks would yield an overall increase of approximately 16 percent. If so, then legal reasoning tasks would achieve an accuracy of 95.7 percent. Because of potential gains to attorney productivity, firms like Lexis Nexis, Westlaw, and Harvey possess the incentives to invest heavily in the construction of tasks and prompts. It seems likely that a capable language model specialized in law will arrive sooner rather than later.

### B. The Computational Future

Once a fully mature LLM is widely adopted in law, lawyers will primarily add value by focusing on non-routine litigation and transactions. As already emphasized, competitive forces will relegate routine legal services to machines with minimal supervision by lawyers. Non-routine tasks, on the other hand, will benefit from more substantial human input. Judges and legislators give society direction when they develop law.[240] Machines can provide suggestions and ideas, but the selection of values is ultimately a human endeavor for the foreseeable future.[241] While language

---

[235] GPT-3.5 and GPT-4 were released, respectively, in November 2022 and March 2023. Kristi Hines, *History of ChatGPT: A Timeline of Developments*, SEARCH ENGINE J. (Jun. 4, 2023), https://www.searchenginejournal.com/history-of-chatgpt-timeline/488370/.

[236] Again, this result is consistent with Nay's study. *See* Nay, *supra* note 224 at *21-22.

[237] *See* Brown, *supra* note 43 at 9.

[238] *See* Ron Miller, *Sam Altman: Size of LLMs Won't Matter as Much Moving Forward*, TECHCRUNCH (Apr. 14, 2023), https://techcrunch.com/2023/04/14/sam-altman-size-of-llms-wont-matter-as-much-moving-forward/ (quoting OpenAI CEO Sam Altman: "I think we're at the end of the era where it's gonna be these giant models, and we'll make them better in other ways.").

[239] *See* Jason Wei, et al., *Chain-of-Thought Prompting Elicits Reasoning in Large Language Models*, 36th Conference on Neural Information Processing Systems 5 Fig. 4 (NeurIPS 2022) (documenting accuracy increases as high as 40 percent when providing ChatGPT with a logically connected series of prompts (as opposed to a single prompt).

[240] *See* GILMORE, *SUPRA* note 67 at 91-93 (describing "judicial activism" as such).

[241] Even if humans were to cede goal-setting and lawmaking to machines, humans could always

models can suggest social goals and provide reasoned argument for their adoption, humans are nowhere near ceding control of their lives to the suggestions of machines.[242] This residual control, exercised by humans, is consequential for the mix of labor between lawyers and AI. A language model might suggest arguments and ideas for new precedent or reversals, but a judge must decide to follow them. Before a judge decides to follow a model's suggestions, a lawyer will want to filter and temper them, provide some of input of her own, and generally combine forces with the model in order to achieve a victory on behalf of a client.[243] This is the likely prerequisite for the model to take part in the discussion in the first place.[244] Thus, while language models will complete more and more of the lawyer's routine work product, advocating for new rules and disrupting older ones will more firmly remain in the hands of the lawyer.

Other possibilities are certainly conceivable.[245] We may one day turn over lawmaking to machines similar to how we allow a constitution to chart a broad course for future lawmaking.[246] After all, the legislature consistently scores very low in public opinion polls, and judiciaries in other societies have been stripped of their powers in response to social upheavals.[247] Nonetheless, the assumption that humans

---

[241] revolt. The permanent possibility of revolt implies that social authority will always rest with people. For further discussion on this point, see Fagan, *supra* note 158 at 69-70.

[242] This is not because of a propensity toward error. It is because language models are currently unable to morally reason. *See* Chomsky, Roberts & Wattumull, *supra* note 1.

[243] In addition to maintaining control over goal setting, combinations of human and AI lawyers will likely outperform AIs working alone. Anecdotal evidence from chess supports this claim. A powerful chess algorithm can beat the best human, but when a human is paired with a chess algorithm as a team, the combination of human plus an AI can often defeat an AI opponent. The human exercises meta-rationality, that is, he knows when the machine judgment is better than his own. However, on occasion, he overrules the machine when he sees what he believes is a potentially better move. *See* COWEN, *SUPRA* note at 77-82 (describing how combinations of humans and AI often triumph over AI alone through exercise of meta-rationality).

[244] This does not mean that the use of an artificial judge is unlikely—only those instances when a decision involves setting a social goal or preference. *See* Fagan & Levmore, *supra* note 67 at 7 (noting that humans will likely set social goals by assigning weights to "efficiency, wealth distribution, [and] short- and long-term climate change," and then ask AI to help achieve those goals).

[245] For instance, an administrative bureaucracy could become all-encompassing and make value judgments for people on the basis of contingent rules. *See, e.g.*, Sofia Ranchordas, *Empathy in the Digital Administrative State*, DUKE L. J. *37 (2023) (describing automated empathy as the ability of machines to compute psycho-physiological states and respond accordingly with administrative decisions).

[246] On the parallels between ceding control to constitutions and machines, see Fagan, *supra* note 158 at 69-70.

[247] *See* Krista Loose, Talking About Congress: The Limited Effect of Congressional Advertising on Congressional Approval, MIT Political Science Department Research Paper No. 2014-30 at 1 (noting consistent low approval). On the severely truncated power of the judiciary in response to revolution in France, see, e.g., THOMAS CARLYLE, FRENCH REVOLUTION: A HISTORY (VOLUME III) 83

will demand residual control over precedent and its underlying social goals seems a reasonable one to make. Even a fully matured AI that makes few mistakes may pursue goals contrary to human preferences and values if given full autonomy to set social policies and create law.[248] By maintaining control of lawmaking, society ensures that its rules are aligned with its preferences. And if judges and lawmakers are making law and choosing the direction for society to take in the process, they will, in all likelihood, want to hear from lawyers, even if those lawyers and their ideas are assisted and advised, in part, by sophisticated machines.

Consider the lawyer's role in the creation and disruption of precedent. When societal and technological innovation provoke reversals of prior precedent, or stimulate questions of first impression, social goals and preferences must be calibrated or set anew. For instance, once automated vehicles are the norm, should accidents be governed by strict liability or negligence, or perhaps, something altogether different like "network liability?"[249] Or once access to Zoom and similar remote conferencing apps is sufficiently widespread, should disabled employees have a right to work online from home?[250] Who should bear the risk of novel smart contract errors—the parties, or should liability fall to programmers and coders in certain scenarios?[251] Each of these questions, and others like them, involve tradeoffs and implicate questions of social judgment. If a community values greater safety to the convenience of automated cars, it will impose greater precautionary standards on manufacturers. If it wishes to expand work-from-home rights, it can experiment with different rules in order to find the optimal regime that balances workers' rights with productivity. Innovations in technology and social organization routinely lead to the creation of new rules and provoke calibrations of existing ones. In each instance of change, society is faced with a choice. Should it maintain status quo law, or select a different path toward a different future? Goals and preferences must be chosen. Value judgments must be made.

---

(2d ed. 1842) (documenting a "completely … shattered Social Constitution … from King down to Parish Constable, all Authorities, Magistrates, [and] Judges").

[248] *See* Chomsky, Roberts & Wattubull, *supra* note 1.

[249] *See* Mark Geistfeld, *A Roadmap for Autonomous Vehicles: State Tort Liability, Automobile Insurance, and Federal Safety Regulation*, 105 CA. L. REV. 1611, 1611 (2017) (asserting that liability for autonomous vehicles, in contrast to standard notions of tort liability, could be based upon the operation of a networked fleet). *See also* OMRI BEN-SHAHAR & ARIEL PORAT, PERSONALIZED LAW: DIFFERENT RULES FOR DIFFERENT PEOPLE 172-73 (2021) (noting that coordination and precautionary safety mechanisms of autonomous driving can be addressed earlier in time at the "programming" level).

[250] *See* D'Andra Shu, *Remote Work Disability Accommodations in the Post-Pandemic Workplace: The Need for Evidence-Driven Analysis*, 95 TEMPLE L. REV. 201 (2023) (noting the possibility).

[251] *See* Daniel Nathan, *A Foreboding View of Smart Contract Developer Liability*, ORRICK (last visited Nov. 29, 2023) (noting that Commissioner Quintenz of the Commodity Futures Trading Commission expressed the possibility of programmer liability based on foreseeability of harm).

Today this work falls to judges and legislators who are assisted by lawyers. Lawyers will, in all likelihood, continue to play a significant role. Their ability to empathize with the winners and losers from a policy decision provides them with a nuanced perspective that a machine could only capture if it were able to compute "the world though the other person's eyes."[252] Even if such computation were possible, the resultant empathy would be computational, and by convention, less empathetic. In short, it would take significant societal change to elevate the programmed compassion and understanding of machines over that of the lawyer. Until then, lawyers will be called upon to help make social value judgments.

## Conclusion[253]

The advent of Large Language Models (LLMs) like ChatGPT heralds a transformative era in the legal profession, one that is poised to reshape the landscape of legal practice and law firm organization significantly. This Article has explored the multifaceted impact of LLMs, emphasizing both the imminent and long-term changes they are likely to bring about in the legal industry.

In the short term, LLMs will stimulate new legal work, primarily by introducing novel legal challenges and questions, particularly around data rights and the implications of AI-generated content. This phase represents a transitional period where lawyers will navigate and shape the legal frameworks governing the use of these advanced technologies. As we look further into the future, the potential for LLMs to streamline and automate routine legal tasks will emerge as a pivotal factor in reshaping standard legal practice. This efficiency gain, while beneficial in terms of productivity, may lead to a significant reduction in the need for legal personnel, particularly in larger law firms whose offerings scale. Such a shift could not only alter employment patterns within the legal sector but also drive a wave of consolidation in the industry, as law firms increasingly compete on the basis of technological prowess rather than solely on legal expertise.

Despite these substantial shifts, the role of human lawyers will remain indispensable, especially in handling complex, non-routine legal tasks that require nuanced judgment, empathy, and an understanding of human values and ethics. The intersection of AI and human expertise will likely define the new frontier of legal practice, where lawyers leverage AI for efficiency while contributing their irreplaceable human judgment and empathetic understanding in shaping and applying the law. Moreover, the evolving landscape suggests that data quality and processing capabilities will become crucial competitive edges. Law firms that can harness superior data and more efficient processing will likely lead the field, potentially ushering in greater market consolidation and a redefinition of what constitutes a leading law firm in the digital age.

---

[252] Minter Dial, Heartificial Empathy: Putting Heart into Business and Artificial Intelligence (2018).

[253] This Conclusion was written with the help of ChatGPT-4. Specifically, I pasted the entire draft of the Article (without the Conclusion) into ChatGPT-4's input window and asked the model to draft a conclusion in the style of the author. It returned an excessively flamboyant summary, almost non-academic in character, which prompted me to ask the model to tone it down. Specifically, I instructed the model to re-draft a conclusion in a "less bombastic" style. That output is presented here, largely intact, as I carried out just a handful of edits to ChatGPT's suggestion. I will leave any inferences for the reader to draw.

In conclusion, while LLMs promise to transform many aspects of legal practice, their integration into the legal profession will be a complex, many-sided process. It necessitates a reevaluation of the lawyer's role, the recalibration of legal education and training, and a thoughtful approach to the ethical and practical implications of AI in law. As the legal profession stands at this crossroads, it faces a dual challenge: leveraging technological advancements for efficiency while preserving the essential human elements at the heart of legal practice.

\* \* \*